\documentclass[onecolumn,12pt,journal,draftclsnofoot]{IEEEtran}
\usepackage[latin9]{inputenc}
\usepackage{float}
\usepackage{multirow}
\usepackage{amsmath}
\usepackage{amssymb}
\usepackage{graphicx}

\makeatletter

\providecommand{\tabularnewline}{\\}
\floatstyle{ruled}
\newfloat{algorithm}{tbp}{loa}
\providecommand{\algorithmname}{Algorithm}
\floatname{algorithm}{\protect\algorithmname}

\usepackage{amsfonts}
\usepackage{mathrsfs}
\usepackage{mathrsfs}\usepackage[noblocks]{authblk}
\usepackage[compress]{cite}
\usepackage{bm}
\usepackage{algorithm}
\usepackage{algorithmic}
\usepackage{epsfig}\usepackage{graphics}\usepackage{subfigure}\usepackage{epsfig}\usepackage{epstopdf}
\usepackage{graphics}\usepackage{subfigure}\usepackage{upgreek}\usepackage{caption}\usepackage{theorem}
\usepackage{breqn}
\usepackage{multicol}
\usepackage{eufrak}
\usepackage{eucal}
\usepackage{array}
\usepackage{cases}
\usepackage[compress]{cite}
\usepackage{algorithm}
\usepackage{algorithmic}
\usepackage{empheq}
\allowdisplaybreaks[4]

\newtheorem{lemma}{Lemma}

\theoremheaderfont{\normalfont\bfseries}

\makeatother

\begin{document}
\title{Channel Estimation for RIS-Aided Multiuser Millimeter-Wave Systems}
\author{Gui~Zhou, Cunhua~Pan, Hong~Ren, Petar~Popovski, \textit{IEEE Fellow},
A. Lee Swindlehurst, \textit{IEEE Fellow}
 \thanks{(Corresponding author: Cunhua Pan)

G. Zhou and C. Pan are with the School of Electronic Engineering and
Computer Science at Queen Mary University of London, London E1 4NS,
U.K. (e-mail: g.zhou, c.pan@qmul.ac.uk). H. Ren is with the National
Mobile Communications Research Laboratory, Southeast University, Nanjing
210096, China. (hren@seu.edu.cn). Petar Popovski is with the Department
of Electronic Systems, Aalborg University, 9220 Aalborg, Denmark (e-mail:
petarp@es.aau.dk). A. L. Swindlehurst is with the Center for Pervasive
Communications and Computing, University of California, Irvine, CA
92697, USA (e-mail: swindle@uci.edu).}}
\maketitle
\begin{abstract}
A reconfigurable intelligent surface (RIS) is a promising device that
can reconfigure the electromagnetic propagation environment through
adjustment of the phase shifts of its multiple reflecting elements.
However, channel estimation in RIS-aided multiuser multiple-input
single-output (MU-MISO) wireless communication systems is challenging
due to the passive nature of the RIS and the large number of reflecting
elements that can lead to high channel estimation overhead. To address
this issue, we propose a novel cascaded channel estimation strategy
with low pilot overhead by exploiting the sparsity and the correlation
of multiuser cascaded channels in millimeter-wave  MISO systems. Based
on the fact that the phsical positions of the BS, the RIS and users
do not appreciably change over multiple consecutive channel coherence
blocks, we first estimate the full channel state information (CSI)
including all the angle and gain information in the first coherence
block, and then only re-estimate the channel gains in the remaining
coherence blocks with much lower pilot overhead. In the first coherence
block, we propose a two-phase channel estimation method, in which
the cascaded channel of one typical user is estimated in Phase I based
on the linear correlation among cascaded paths, while the cascaded
channels of other users are estimated in Phase II by utilizing the
reparameterized CSI of the common base station (BS)-RIS channel obtained
in Phase I. The minimum pilot overhead is much less than the existing
works. Simulation results show that the performance of the proposed
method outperforms existing methods in terms of the estimation accuracy
when using the same amount of pilot overhead. 
\end{abstract}


\section{Introduction}

A reconfigurable intelligent surface (RIS) can enhance the coverage
and capacity of wireless communication systems with relatively low
hardware cost and energy consumption \cite{Marco-5,Pan-mag,Pan2019intelleget,Pan2019multicell,shanpu}.
An RIS is typically composed of a large number of passive elements,
which can assist the wireless communication by reconfiguring the electromagnetic
propagation environment between a transmitter and receiver. The performance
gain provided by the RIS relies heavily on the accuracy of the channel
state information (CSI). However, it is challenging to acquire the
CSI since the reflecting elements at the RIS are passive devices lacking
the ability of transmitting, receiving and processing pilot signals.

It is observed that the CSI of the cascaded base station (BS)-IRS-user
channel, which is the product of the BS-IRS channel and the IRS-user
channel, is sufficient for the transmission design \cite{Gui2019IRS,Xianghao2009}.
As a result, most of the existing contributions have focused on cascaded
channel estimation \cite{LS-mvue,2019beixiong,shuguang-IRS,ris-omp-1,2021Jiguang,ris-omp-2,ris-omp-3}.
Specifically, consider a system containing a BS with $N$ antennas,
$K$ single-antenna users, and one IRS with $M$ reflecting elements.
The authors in \cite{LS-mvue} proposed a least-squares (LS)-based
estimation method to obtain the cascaded channel estimator which is
unbiased for single-user multiple-input single-output (SU-MISO) systems.
However, the pilot overhead of the LS-based estimation method is prohibitively
high and scales with $M$, which can be quite large. To reduce the
pilot overhead, \cite{2019beixiong} divided the elements of the RIS
into $P$ subgroups, and proposed a transmission protocol to successively
excecute channel estimation and phase shift optimization with a pilot
overhead of $P$. By exploiting the common BS-RIS channel and the
linear correlation among the RIS-user channels in multiuser multiple-input
single-output (MU-MISO) systems, the authors in \cite{shuguang-IRS}
further proposed a channel estimation strategy whose pilot overhead
is inversely proportional to the number of the antennas at the BS:
$M+\max(K-1,K\left\lceil \frac{(K-1)M}{N}\right\rceil )$. The estimation
method in \cite{shuguang-IRS} requires low pilot overhead in a rich
scattering communication scenario where the cascaded channel is full
rank, but this method is not applicable in millimeter-wave (mmWave)
 MISO communication systems where the channel is rank-deficient due
to the spatial sparsity \cite{mmWave-channel}.

To address this issue, the authors in \cite{ris-omp-1,2021Jiguang,ris-omp-2,ris-omp-3}
exploited the sparsity of the cascaded channel matrix in mmWave communication
systems and proposed compressed sensing (CS)-based channel estimation
methods with low pilot overhead. In particular, \cite{ris-omp-1}
directly constructed a sparse signal recovery problem for cascaded
channel estimation, but ignored the common parameters of the cascaded
channel in SU-MISO systems, which leads to high power leakage. Thus,
the adopted on-grid CS method has high false alarm probability and
high estimation error. In order to suppress the power leakage effect,
the atomic norm minimization method was used in \cite{2021Jiguang}
to estimate the sparse angles and gains. For MU-MISO systems, both
\cite{ris-omp-2} and \cite{ris-omp-3} investigated the double sparse
structure of the cascaded channel and utilized common parameters to
jointly estimate the multiuser cascaded channels with low pilot overhead
and high estimation accuracy. However, these two papers assumed that
the number of BS-RIS channel paths $L$ and the number of RIS-user
channel paths $J$ are known \textit{a priori}, an assumption that
is difficult to achieve in practic. Moreover, the pilot overhead in
\cite{ris-omp-2} is proportional to the quotient of the number of
RIS elements divided by the number of cascaded spatial paths, i.e.,
$K\left\lceil \frac{M}{JL}\right\rceil $, which can be excessively
large in large RIS systems with a large number of reflecting elements.
Therefore, this motivates the development of an efficient channel
estimation strategy to further reduce the pilot overhead, as well
as estimate the sparsity level, or equivalently the number of spatial
paths.

\subsection{Novelty and contributions}

Against the above backdrop, this paper proposes a novel uplink cascaded
channel estimation strategy for RIS-aided multiuser mmWave systems.
The proposed estimation strategy has the following appealing features:
low pilot overhead, low computational complexity, and estimation of
the sparsity level (number of spatial paths) of the cascaded channel.
These appealing features are achieved based on the following three
typical properties:

\textbf{Property 1: }The physical positions of the BS and the RIS
change much more slowly than the individual channel coefficients \cite{mmWave-channel}.
Therefore, it is reasonable to assume that the angles-of-arrival (AoAs)
at the BS, and the AoAs and angles-of-departure (AoDs) at the RIS
remain unchanged over multiple channel coherence blocks. If the angle
information is estimated in the first channel coherence block, only
the cascaded channel gains need to be re-estimated in the subsequent
channel coherence blocks. This can greatly reduce the pilot overhead
and computational complexity of channel estimation in later blocks,
since only a few parameters need to be estimated.

\textbf{Property 2: }The $JL$ cascaded paths are the combination
of $J+L$ independent spatial paths. This means that there is a linear
correlation among the $JL$ cascaded paths, which motivates the direct
estimation of the $J+L$ sparse paths, rather than the $JL$ cascaded
sparse paths. Note that the existing contributions in \cite{ris-omp-1,2021Jiguang,ris-omp-2,ris-omp-3}
estimate $JL$ cascaded sparse paths.

\textbf{Property 3: }All users share a common BS-RIS channel. Based
on this property, \cite{ris-omp-2,ris-omp-3} exploited the common
AoA information of the BS-RIS channel to simplify the multiuser channel
estimation and reduce pilot overhead. In this work, we exploit the
AoA, AoD and gain information of the common BS-RIS channel to construct
a reparameterized common BS-RIS channel, which enables us to develop
a new multiuser channel estimation method with less pilot overhead.

Based on the above discussion, the main contributions of this work
are summarized as follows: 
\begin{itemize}
\item We propose a novel uplink channel estimation protocol for time division
duplex (TDD) RIS-aided multiuser mmWave communication systems, as
depicted in Fig. \ref{channel estimation}. Based on \textbf{Property
1}, we assume that the angle parameters of the CSI remain constant
over multiple channel coherence blocks, while the channel gains vary
from block to block. In the first coherence block, we estimate the
full CSI, including all the angle information and the channel gains.
Given the estimated angle information, only the channel gains need
to be estimated in the remaining coherence blocks, which can be achieved
using a simple LS method with a low overhead of $JK$ pilots. Moreover,
the training phase shift matrices are optimized to minimize the mutual
coherence of the equivalent dictionary for better estimation performance. 
\item In the first coherence block, we propose a two-phase channel estimation
method that makes use of \textbf{Property 2} and \textbf{Property
3}. In particular, in Phase I, a typical user sends a sequence of
pilots to the BS for cascaded channel estimation. The required theoretical
minimum pilot overhead can be made as low as $8J-2$ by exploiting
the linear correlation among the cascaded paths based on \textbf{Property
2}. Based on \textbf{Property 3}, we extract the reparameterized CSI
of the common BS-RIS channel from Phase I, which can help reduce the
pilot overhead for estimation of the CSI of other users.. In Phase
II, the other users successively transmit pilots to the BS for channel
estimation. With knowledge of the reparameterized common BS-RIS channel,
the minimum required pilot overhead can be reduced to $(K-1)\left\lceil (8J-2)/L\right\rceil $.
Therefore, the minimum pilot overhead in the first coherence block
is $8J-2+(K-1)\left\lceil (8J-2)/L\right\rceil $. 
\item We demonstrate through numerical results that the proposed cascaded
channel estimation strategy outperforms the existing orthogonal matching
pursuit (OMP)-based channel estimation algorithm in terms of mean
squared error (MSE), the pilot overhead and the computational complexity.
Moreover, the MSE performance of the proposed estimation algorithm
is close to the performance lower bound at low SNR. 
\end{itemize}
\,\,\,\,\,\,\,The remainder of this paper is organized as follows.
Section II introduces the system model and the cascaded channel sparsity
model. The cascaded channel estimation strategy is investigated in
Section III. Training phase shift matrices are optimized in Sections
IV. Section V compares the pilot overhead and algorithm complexity
between the proposed algorithm and existing algorithms. Finally, Sections
VI and VII report the numerical results and conclusions, respectively.

\noindent \textbf{Notations:} The following mathematical notations
and symbols are used throughout this paper. Vectors and matrices are
denoted by boldface lowercase letters and boldface uppercase letters,
respectively. The symbols $\mathbf{X}^{*}$, $\mathbf{X}^{\mathrm{T}}$,
$\mathbf{X}^{\mathrm{H}}$, and $||\mathbf{X}||_{F}$ denote the conjugate,
transpose, Hermitian (conjugate transpose), Frobenius norm of matrix
$\mathbf{X}$, respectively. The symbol $||\mathbf{x}||_{2}$ denotes
2-norm of vector $\mathbf{x}$. The symbols $\mathrm{Tr}\{\cdot\}$,
$\mathrm{Re}\{\cdot\}$, $|\cdot|$, and $\angle\left(\cdot\right)$
denote the trace, real part, modulus, and angle of a complex number,
respectively. $\mathrm{Diag}(\mathbf{x})$ is a diagonal matrix with
the entries of vector $\mathbf{x}$ on its main diagonal. $[\mathbf{x}]_{m}$
denotes the $m$-th element of the vector $\mathbf{x}$, and $[\mathbf{X}]_{m,n}$
denotes the $(m,n)$-th element of the matrix $\mathbf{X}$. $\mathbf{X}_{(:,n)}$
and $\mathbf{X}_{(m,:)}$ denote the $n$-th column and the $m$-th
row of matrix $\mathbf{X}$. The Kronecker and Khatri-Rao products
between two matrices $\mathbf{X}$ and $\mathbf{Y}$ are denoted by
$\mathbf{X}\otimes\mathbf{Y}$ and $\mathbf{X}\odot\mathbf{Y}$, respectively.
Additionally, the symbol $\mathbb{C}$ denotes complex field, $\mathbb{R}$
represents real field, and $\mathrm{i}\triangleq\sqrt{-1}$ is the
imaginary unit. The inner product $\left\langle \bullet,\bullet\right\rangle :\mathbb{C}^{M\times N}\times\mathbb{C}^{M\times N}\rightarrow\mathbb{R}$
is defined as $\left\langle \mathbf{X},\mathbf{Y}\right\rangle =\mathbb{R}\{\mathrm{Tr}\{\mathbf{X}^{\mathrm{H}}\mathbf{Y}\}\}.$
$\left\lceil \right\rceil $ rounds up to the nearest integer, and
$\left\lceil \right\rfloor $ rounds to the closest integer.

\section{System and Channel Model}

\subsection{Signal Model}

We consider a narrow-band TDD mmWave massive MISO system where $K$
single-antenna users communicate with an $N$-antenna BS. To enhance
the spatial diversity and improve communication performance, an RIS
equipped with $M$ passive reflecting elements, each of which can
be dynamically adjusted for electromagnetic wave reconstruction between
the BS and users, is deployed.

In this paper, we consider quasi-static block-fading channels, where
each channel remains approximately constant in a channel coherence
block with $B$ time slots. Due to channel reciprocity, the CSI of
the downlink channel can be obtained by estimating the CSI of the
uplink channel. We assume that $\tau$ time slots of each coherence
block are used for uplink channel estimation and the remaining $B-\tau$
time slots for downlink data transmission. Here, we assume that the
direct channels between the BS and users are blocked\footnote{If the direct channels between the BS and users are available, then
the CSI of the direct channels can be obtained with the RIS turned
off \cite{shuguang-IRS}.}. Therefore, we only focus on the uplink channel estimation of the
user-RIS links and the RIS-BS link.

Let $\mathbf{h}_{k}\in\mathbb{C}^{M\times1}$ denote the channel from
user $k$ to the RIS and $\mathbf{H}\in\mathbb{C}^{N\times M}$ denote
the channel from the RIS to the BS. Moreover, denote by $\mathbf{e}_{t}\in\mathbb{C}^{M\times1}$
the phase shift vector of the RIS at time slot $t$ in the considered
coherence block, which satisfies $|[\mathbf{e}_{t}]_{m}|^{2}=1$ for
$1\leq m\leq M$. Define set $\mathcal{K}=\{1,\ldots,K\}$. Here,
we assume that the users transmit pilot sequences of length $\tau_{k}$
one by one for channel estimation. The received signal from user $k$
at the BS after removing the impact of the direct channel at time
slot $t$, $1\leq t\leq\tau_{k}$, can be expressed as 
\begin{align}
\mathbf{y}_{k}(t) & =\mathbf{H}\mathrm{Diag}(\mathbf{e}_{t})\mathbf{h}_{k}\sqrt{p}s_{k}(t)+\mathbf{n}_{k}(t),\forall k\in\mathcal{K},\label{eq:1}
\end{align}
where $s_{k}(t)$ and $\mathbf{n}_{k}(t)\in\mathbb{C}^{N\times1}\sim\mathcal{CN}(0,\delta^{2}\mathbf{I})$
denote the transmitted pilot signal of the $k$-th user and additive
white Gaussian noise (AWGN) with power $\delta^{2}$ at the BS at
time slot $t$, respectively. The quantity $p$ denotes the transmit
power of each user, which for simplicity is assumed here to be the
same for all users.

Equation (\ref{eq:1}) can be rewritten as 
\begin{align}
\mathbf{y}_{k}(t) & =\mathbf{H}\mathrm{Diag}(\mathbf{h}_{k})\mathbf{e}_{t}\sqrt{p}s_{k}(t)+\mathbf{n}_{k}(t),\forall k\in\mathcal{K}.\label{eq:2}
\end{align}
This indicates that joint design of the active beamforming at the
BS and the passive reflecting beamforming at the RIS depends on the
cascaded user-RIS-BS channels \cite{Gui2019IRS,Xianghao2009}: 
\begin{equation}
\mathbf{G}_{k}=\mathbf{H}\mathrm{Diag}(\mathbf{h}_{k})\in\mathbb{C}^{N\times M},\forall k\in\mathcal{K}.\label{eq:3}
\end{equation}
Our work focuses on estimation of the cascaded channels in (\ref{eq:3}).

Consider user $k$ who transmits $\tau_{k}$ pilot symbols to the
BS. For simplicity, we assume that the pilot symbols satisfy $s_{k}(t)=1,1\leq t\leq\tau_{k}$.
The measurement matrix $\mathbf{Y}_{k}=\left[\mathbf{y}_{k}(1),\ldots,\mathbf{y}_{k}(\tau_{k})\right]\in\mathbb{C}^{N\times\tau_{k}}$
received at the BS during user $k$'s pilot transmission is expressed
as 
\begin{equation}
\mathbf{Y}_{k}=\sqrt{p}\mathbf{G}_{k}\mathbf{E}_{k}+\mathbf{N}_{k}\in\mathbb{C}^{N\times\tau_{k}},\label{eq:m-y-2}
\end{equation}
where \begin{subequations} \label{hc-a-1} 
\begin{align}
\mathbf{E}_{k} & =\left[\mathbf{e}_{1},\ldots,\mathbf{e}_{\tau_{k}}\right]\in\mathbb{C}^{M\times\tau_{k}},\label{cd}\\
\mathbf{N}_{k} & =\left[\mathbf{n}_{k}(1),\ldots,\mathbf{n}_{k}(\tau_{k})\right]\in\mathbb{C}^{N\times\tau_{k}}.\label{cd4}
\end{align}
\end{subequations}According to \cite{LS-mvue}, the LS estimator
\begin{equation}
\mathbf{G}_{k}^{\mathrm{LS}}=\frac{1}{\sqrt{p}}\mathbf{Y}_{k}\mathbf{E}_{k}^{\mathrm{H}}(\mathbf{E}_{k}\mathbf{E}_{k}^{\mathrm{H}})^{-1}\label{ls}
\end{equation}
of $\mathbf{G}_{k}$ is unbiased when the design of the phase shift
matrix $\mathbf{E}_{k}$ is chosen in a particular way. However, the
required pilot overhead $\tau_{k}\geq M$ for each user is unacceptable
due to the fact that the RIS is generally equipped with a large number
of elements. Therefore, it is of interest to investigate more efficient
channel estimation strategies that reduce the pilot overhead by exploiting
the sparsity of the mmWave massive MISO channel.

\subsection{Cascaded Channel Sparsity Model}

It is assumed that both BS and RIS are equipped with a uniform linear
array (ULA) with antenna spacing $d_{\mathrm{BS}}$ and $d_{\mathrm{RIS}}$,
respectively. Applying the geometric channel model typically used
for mmWave systems \cite{mmWave-channel}, channels $\mathbf{H}$
and $\mathbf{h}_{k}$ are modeled as 
\begin{align}
\mathbf{H} & =\sum_{l=1}^{L}\alpha_{l}\mathbf{a}_{N}\left(\psi_{l}\right)\mathbf{a}_{M}^{\mathrm{H}}\left(\omega_{l}\right),\label{eq:H1}\\
\mathbf{h} & _{k}=\sum_{j=1}^{J_{k}}\beta_{k,j}\mathbf{a}_{M}\left(\varphi_{k,j}\right),\forall k\in\mathcal{K},\label{eq:H}
\end{align}
where $L$ and $J_{k}$ denote the number of propagation paths between
the BS and the RIS and between the RIS and user $k$, respectively.
The complex gains of the $l$-th path in the BS-RIS channel and the
$j$-th path in the RIS-user-$k$ channel are represented by $\alpha_{l}$
and $\beta_{k,j}$, respectively. Denote by $\mathbf{a}_{X}(x)\in\mathbb{C}^{X\times1}$
the array steering vector, i.e., 
\begin{align*}
\mathbf{a}_{X}(x) & =[1,e^{-\mathrm{i}2\pi x},\ldots,e^{-\mathrm{i}2\pi(X-1)x}]^{\mathrm{T}},
\end{align*}
where $X\in\{M,N\}$ and $x\in\{\omega_{l},\psi_{l},\varphi_{k,j}\}$.
$\omega_{l}=\frac{d_{\mathrm{RIS}}}{\lambda_{c}}\cos(\theta_{l})$,
$\psi_{l}=\frac{d_{\mathrm{BS}}}{\lambda_{c}}\cos(\phi_{l})$, and
$\varphi_{k,j}=\frac{d_{\mathrm{RIS}}}{\lambda_{c}}\cos(\vartheta_{k,j})$
are the directional cosines, where $\theta_{l}$ and $\phi_{l}$ respectively
denote the AoD and AoA of the $l$-th spatial path from RIS to BS,
and $\vartheta_{k,j}$ is the AoA of the $j$-th spatial path from
user $k$ to the RIS. $\lambda_{c}$ is the carrier wavelength. It
should be emphasized here that the channel gains $\alpha_{l}$ and
$\beta_{k,j}$ change at each channel coherence block, while the angles
$\{\theta_{l},\phi_{l},\vartheta_{k,j}\}$ vary much more slowly than
the channel gains, and generally remain invariant during multiple
channel coherence blocks.

From (\ref{eq:H1}) and (\ref{eq:H}), the geometric model of the
cascaded channels in (\ref{eq:3}) is formulated as 
\begin{align}
\mathbf{G}_{k} & =\sum_{l=1}^{L}\sum_{j=1}^{J_{k}}\alpha_{l}\beta_{k,j}\mathbf{a}_{N}(\psi_{l})\mathbf{a}_{M}^{\mathrm{H}}(\omega_{l}-\varphi_{k,j}),\forall k\in\mathcal{K}.\label{eq:G2}
\end{align}
Note that $\mathbf{a}_{M}(\omega_{l}-\varphi_{k,j})$ is the steering
vector of the $jl$-th cascaded subpath of user $k$, and the corresponding
term $\cos(\theta_{l})-\cos(\vartheta_{k,j})$ is called as the cosine
of the cascaded AoD for the $jl$-th cascaded subpath from user $k$.

The channel model in (\ref{eq:G2}) illustrates the low rank property
and the spatial correlation characteristics of RIS-aided mmWave  system.
Thus, CS-based sparse cascaded channel estimation methods are widely
used based on the expression in (\ref{eq:G2}) \cite{ris-omp-1,ris-omp-2,ris-omp-3}.
In particular, (\ref{eq:G2}) is approximated using the virtual angular
domain (VAD) representation, i.e., 
\begin{equation}
\mathbf{G}_{k}=\mathbf{A}_{R}\mathbf{X}_{k}\mathbf{A}_{T}^{\mathrm{H}},\label{ds}
\end{equation}
where dictionary matrices $\{\mathbf{A}_{R},\mathbf{A}_{T}\}$ can
be drawn from the array steering vectors \cite{ris-omp-2,ris-omp-1}
or from the DFT matrix \cite{ris-omp-3}. The matrix $\mathbf{X}_{k}$
is the angular domain cascaded channel matrix containing $J_{k}L$
complex channel gains, which exhibits sparsity. The CS-based estimation
methods in \cite{ris-omp-1,ris-omp-2,ris-omp-3} need to estimate
$L$ AoAs, $J_{k}L$ cascaded AoD cosines, and $J_{k}L$ cascaded
complex channel gains. The number of parameters to be estimated in
\cite{ris-omp-1,ris-omp-2,ris-omp-3} is much less than in the LS
estimator of \cite{LS-mvue}, since the number of spatial paths is
usually much less than the number of antennas, i.e., $J_{k}L\ll N$
and $J_{k}L\ll M$. However, we can further reduce the number of parameters
to be estimated by exploiting the structure of the cascaded channel.

Specifically, (\ref{eq:H1}) is reformulated as 
\begin{align}
\mathbf{H} & =\mathbf{A}_{N}\boldsymbol{\Lambda}\mathbf{A}_{M}^{\mathrm{H}},\label{eq:H1-1}
\end{align}
where \begin{subequations} \label{Hv} 
\begin{align}
\mathbf{A}_{N} & =[\mathbf{a}_{N}(\psi_{1}),\ldots,\mathbf{a}_{N}(\psi_{L})]\in\mathbb{C}^{N\times L},\label{H1}\\
\boldsymbol{\Lambda} & =\mathrm{Diag}(\alpha_{1},\alpha_{2},\ldots,\alpha_{L})\in\mathbb{C}^{L\times L},\label{H2}\\
\mathbf{A}_{M} & =[\mathbf{a}_{M}(\omega_{1}),\ldots,\mathbf{a}_{M}(\omega_{L})]\in\mathbb{C}^{M\times L}.\label{H3}
\end{align}
\end{subequations}Equation (\ref{eq:H}) is rewritten as 
\begin{align}
\mathbf{h}_{k} & =\mathbf{A}_{M,k}\boldsymbol{\beta}_{k},\forall k\in\mathcal{K},\label{eq:H-1-1}
\end{align}
where \begin{subequations} \label{hv} 
\begin{align}
\mathbf{A}_{M,k} & =[\mathbf{a}_{M}(\varphi_{k,1}),\ldots,\mathbf{a}_{M}(\varphi_{k,J_{k}})]\in\mathbb{C}^{M\times J_{k}},\label{h1}\\
\boldsymbol{\beta}_{k} & =[\beta_{k,1},\ldots,\beta_{k,J_{k}}]^{\mathrm{T}}\in\mathbb{C}^{J_{k}\times1}.\label{h2}
\end{align}
\end{subequations}Hence, (\ref{eq:3}) becomes 
\begin{align}
\mathbf{G}_{k} & =\mathbf{A}_{N}\boldsymbol{\Lambda}\mathbf{A}_{M}^{\mathrm{H}}\mathrm{Diag}\left(\mathbf{A}_{M,k}\boldsymbol{\beta}_{k}\right),\forall k\in\mathcal{K}.\label{eq:G1}
\end{align}

It is observed from (\ref{eq:G1}) that there are actually only $J_{k}+L$
complex gains and $2L+J_{k}$ angles (or directional cosines) that
need to be estimated for each user. In addition, due to the fact that
all the users share the common BS-RIS channel $\mathbf{H}$, they
share the same $L$ complex gains $\{\alpha_{l}\}_{l=1}^{L}$ and
$2L$ angles $\{\theta_{l},\phi_{l}\}_{l=1}^{L}$. Based on this observation,
we develop a novel channel estimation strategy in this work. We remark
that the contributions in \cite{ris-omp-2} and \cite{ris-omp-3}
only take advantage of the information from the common angles $\{\phi_{l}\}_{l=1}^{L}$
and ignore the information from common gains $\{\alpha_{l}\}_{l=1}^{L}$
and common angles $\{\theta_{l}\}_{l=1}^{L}$.

\section{Channel Estimation}

\subsection{Channel Estimation Protocol}

\begin{figure}
\centering \includegraphics[width=9cm,height=2.8cm]{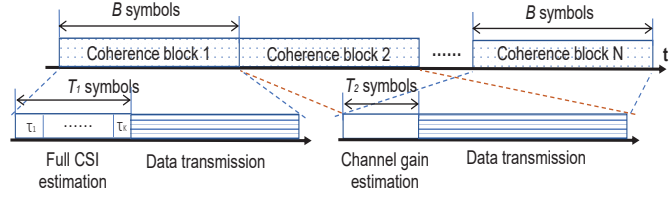}
\caption{Channel estimation protocol and frame structure.}
\label{channel estimation} 
\end{figure}

In this section, we develop a novel uplink channel estimation protocol
by exploiting the sparsity of the RIS-aided mmWave  channel, as shown
in Fig. \ref{channel estimation}.

In most situations, the BS and RIS are in fixed positions, and the
users do not move a significant distance over milliseconds or even
seconds, which corresponds to many channel coherence blocks. Based
on this observation, we assume a model in which the angles remain
unchanged for multiple coherence blocks, while the gains change from
block to block \cite{mmWave-channel}. In the first coherence block,
we estimate the full CSI information, including all the angle information
and the channel gains. We then only need to estimate the channel gains
in the remaining coherence blocks, which can be obtained using a simple
LS method with a significantly smaller set of pilot symbols.

The most difficult aspect of the algorithm is estimation of the full
CSI in the first coherence block. The main idea is explained as follows.
First, a typical user, denoted as user 1 for convenience, sends a
pilot sequence of $\tau_{1}$ symbols to the BS for channel estimation
using CS techniques. With knowledge of the estimated AoAs, cascaded
AoD cosines, and cascaded gains of user 1, we construct a reparameterized
common BS-RIS channel with known CSI, which can be exploited to reduce
the channel estimation overhead associated with users $2$ through
$K$. Then, the remaining users successively transmit pilot symbols
to the BS for channel estimation. Note while the channel estimation
in the first coherence block is time consuming, it will only be performed
once at the start of the transmission.

\subsection{Channel Estimation for User 1 in the First Coherence Block}

In this subsection, we provide the channel estimation method for user
1 with low pilot overhead by exploiting the properties of massive
antenna arrays and the structure of the cascaded channel.

\subsubsection{Estimation of the common AoAs}

\begin{figure*}
\centering \includegraphics[width=7in,height=3in]{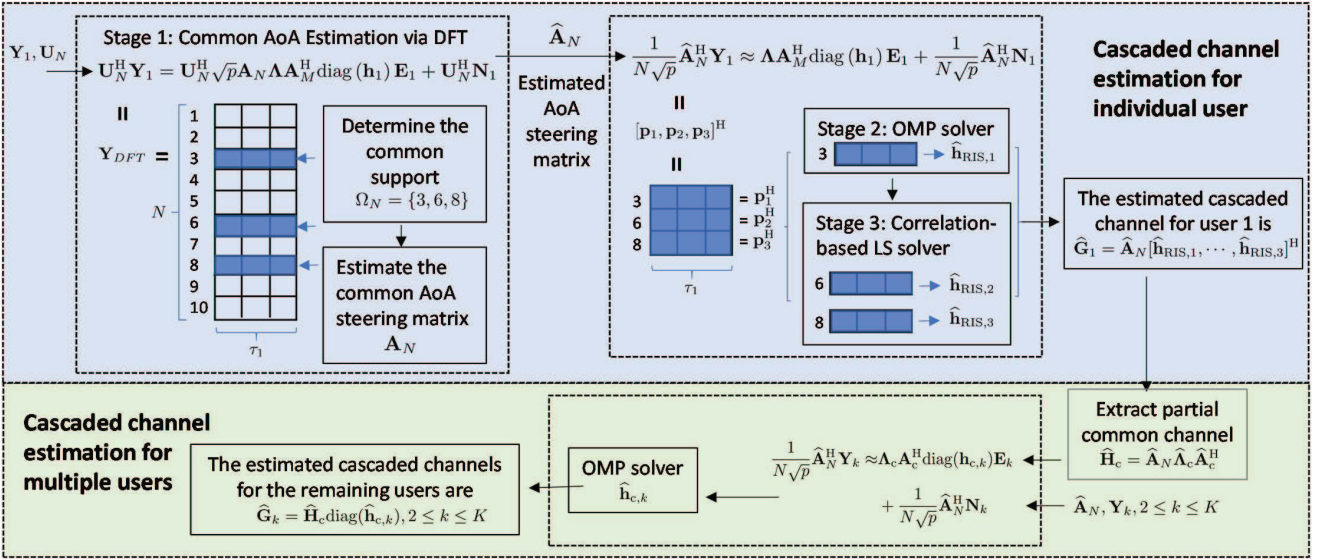} \caption{Cascaded channel estimation strategy for multiple users.}
\label{stage} 
\end{figure*}

Due to the large number of antennas at the BS, the discrete Fourier
transform (DFT) approach can be applied efficiently for AoA estimation
from $\mathbf{Y}_{1}$ in (\ref{eq:m-y-2}). We first present the
asymptotic properties of $\mathbf{A}_{N}$ in the following lemmas,
whose proofs are provided in Appendix \ref{subsec:The-proof-of-3}
and Appendix \ref{subsec:The-proof-of-1}.

\begin{lemma}\label{othogonal} When $N\rightarrow\infty$, the following
property holds 
\begin{equation}
\lim_{N\rightarrow\infty}\frac{1}{N}\mathbf{a}_{N}^{\mathrm{H}}(\psi_{j})\mathbf{a}_{N}(\psi_{i})=\begin{cases}
1 & \psi_{j}=\psi_{i}\\
0 & \textrm{otherwise}
\end{cases},\label{fd-1}
\end{equation}
and $\mathbf{A}_{N}^{\mathrm{H}}\mathbf{A}_{N}=N\mathbf{I}_{L}$,
where $\mathbf{I}_{L}$ is the identity matrix of dimension $L\times L$.
\end{lemma}

\begin{lemma}\label{DFT} When $N\rightarrow\infty$, if the condition
$\frac{d_{\mathrm{BS}}}{\lambda_{c}}\leq1$ holds, then the DFT of
$\mathbf{A}_{N}$, i.e., $\mathbf{U}_{N}^{\mathrm{H}}\mathbf{A}_{N}$,
is a tall sparse matrix with one nonzero element in each column 
\[
\lim_{N\rightarrow\infty}[\mathbf{U}_{N}^{\mathrm{H}}\mathbf{A}_{N}]_{n_{l},l}\neq0,\forall l,
\]
where $\mathbf{U}_{N}$ is the normalized DFT matrix with $(n,m)$-th
entry $[\mathbf{U}_{N}]_{n,m}=\frac{1}{\sqrt{N}}e^{-\mathrm{i}\frac{2\pi}{N}(n-1)(m-1)}$,
and 
\begin{equation}
n_{l}=\begin{cases}
N\psi_{l}+1 & \psi_{l}\in[0,\frac{d_{\mathrm{BS}}}{\lambda_{c}})\\
N+N\psi_{l}+1 & \psi_{l}\in[-\frac{d_{\mathrm{BS}}}{\lambda_{c}},0)
\end{cases}.\label{eq:ss-1}
\end{equation}

Based on Lemma \ref{DFT}, any two nonzero elements are not in the
same row, i.e., $n_{l}\neq n_{i}$ for any $l\neq i$. \end{lemma}

\textbf{Remark 1:} It is observed from (\ref{eq:ss-1}) that when
$\psi_{l}\in[0,\frac{d_{\mathrm{BS}}}{\lambda_{c}})$, the range of
$n_{l}$ is $n_{l}\in\text{[}1,N\frac{d_{\mathrm{BS}}}{\lambda_{c}}+1)$.
When $\psi_{l}\in[-\frac{d_{\mathrm{BS}}}{\lambda_{c}},0)$, we have
$n_{l}\in[N-N\frac{d_{\mathrm{BS}}}{\lambda_{c}}+1,N+1)$. In order
to avoid ambiguous angles where the same $n_{l}$ corresponds to two
AoAs, we much have $N\frac{d_{\mathrm{BS}}}{\lambda_{c}}+1\leq N-N\frac{d_{\mathrm{BS}}}{\lambda_{c}}+1$,
which leads to $d_{\mathrm{BS}}\leq\frac{\lambda_{c}}{2}$. Therefore,
$d_{\mathrm{BS}}$ should generally be restricted to be no larger
than $\lambda_{c}/2$ to avoid AoA ambiguity.

Based on Lemma \ref{DFT}, matrix $\mathbf{U}_{N}^{\mathrm{H}}\mathbf{A}_{N}$
can be regarded as a row sparse matrix with full column rank. Thus,
the DFT of $\mathbf{Y}_{1}$, i.e., $\mathbf{Y}_{DFT}=\mathbf{U}_{N}^{\mathrm{H}}\mathbf{Y}_{1}=\sqrt{p}\mathbf{U}_{N}^{\mathrm{H}}\mathbf{A}_{N}\boldsymbol{\Lambda}\mathbf{A}_{M}^{\mathrm{H}}\mathrm{Diag}\left(\mathbf{h}_{1}\right)\mathbf{E}_{1}+\mathbf{U}_{N}^{\mathrm{H}}\mathbf{N}_{1}$,
is an asymptotic row sparse matrix with $L$ nonzero rows, each corresponding
to one of the AoAs as shown in Fig. \ref{stage}. Based on the above
discussion, $\phi_{l}$ can be immediately estimated from the nonzero
rows of $\mathbf{Y}_{DFT}$. However, $N$ is finite in practice,
and thus $N\psi_{l}$ is usually not an integer. Most of the power
of $\mathbf{Y}_{DFT}$ will be concentrated on the $(\left\lfloor N\psi_{l}\right\rceil +1)$-th
or the $(N+\left\lfloor N\psi_{l}\right\rceil +1)$-th row, while
the remaining power leaks to nearby rows. This is known as the power
leakage effect \cite{feifei-JSAC2017,feifei-TSP2018,feifei-TWC2018,feifei-TSP2019}.
Due to the fact that the resolution of the DFT is $1/N$, there exists
a mismatch between the discrete estimated angle and the real continuous
angle. To improve the angle estimation accuracy, we adopt an angle
rotation operation to compensate for the mismatch of the DFT \cite{feifei-JSAC2017,feifei-TSP2018,feifei-TWC2018}.

The angle rotation matrices are defined as 
\begin{align*}
\boldsymbol{\Phi}_{N}(\triangle\psi_{l}) & =\mathrm{Diag}\{1,e^{\mathrm{i}\triangle\psi_{l}},\ldots,e^{\mathrm{i}(N-1)\triangle\psi_{l}}\},\forall l,
\end{align*}
where $\triangle\psi_{l}\in[-\frac{\pi}{N},\frac{\pi}{N}]$ are the
phase rotation parameters. Then, the angle rotation of $\mathbf{Y}_{1}$
for $\phi_{l}$ is defined as 
\begin{equation}
\mathbf{Y}_{1,l}^{ro}=\boldsymbol{\Phi}_{N}^{\mathrm{H}}(\triangle\psi_{l})\mathbf{Y}_{1}.\label{eq:angle-ro}
\end{equation}
The aim of the angle rotation in (\ref{eq:angle-ro}) is to rotate
$\mathbf{A}_{N}$ in the angle domain such that there is no power
leakage for estimating $\psi_{l}$. For better illustration, we take
$\mathbf{U}_{N}^{\mathrm{H}}\boldsymbol{\Phi}_{N}^{\mathrm{H}}(\triangle\psi_{l})\mathbf{A}_{N}$
as an example, whose $(n,l)$-th element is calculated as 
\begin{align*}
[\mathbf{U}_{N}^{\mathrm{H}}\boldsymbol{\Phi}_{N}^{\mathrm{H}}(\triangle\psi_{l})\mathbf{A}_{N}]_{n,l} & =\sqrt{\frac{1}{N}}\sum_{m=1}^{N}e^{-\mathrm{i}2\pi(m-1)(\psi_{l}+\frac{\triangle\psi_{l}}{2\pi}-\frac{n-1}{N})}.
\end{align*}
It can be readily found that the channel power of $\psi_{l}$ is concentrated
on the $n_{l}$-th row without power leakage when the phase rotation
parameter satisfies 
\begin{align}
\triangle\psi_{l} & =2\pi\left(\frac{n_{l}-1}{N}-\psi_{l}\right).\label{ro1}
\end{align}

For $\mathbf{Y}_{1}$, the optimal phase rotation parameter for $\psi_{l}$
can be found based on a one-dimensional search by solving the following
problem 
\begin{align}
\triangle\psi_{l}=\mathrm{arg}\max_{\triangle\psi\in[-\frac{\pi}{N},\frac{\pi}{N}]} & ||[\mathbf{U}_{N}]_{:,n_{l}}^{\mathrm{H}}\boldsymbol{\Phi}_{N}^{\mathrm{H}}(\triangle\psi)\mathbf{Y}_{1}||^{2}.\label{eq:rotation}
\end{align}

\begin{figure}
\centering \includegraphics[width=3in,height=1.5in]{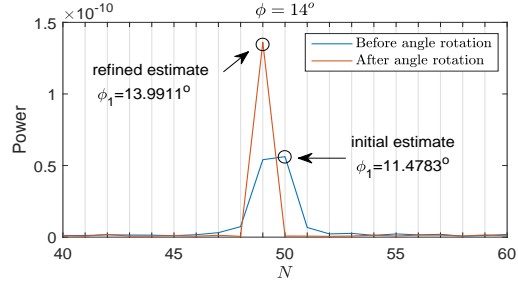} \caption{An example of the row sparse characteristic of $\mathbf{Y}_{DFT}$
and optimal angle rotation, when $L=1$ and $N=M=100$.}
\label{rotation} 
\end{figure}

Fig. \ref{rotation} is an example of the row sparse characteristic
of $\mathbf{Y}_{DFT}$ and the Y-axis is the power of each row of
$\mathbf{Y}_{DFT}$. The cascaded channel of size $N=M=100$ contains
one ($L=1$) path between the BS and the RIS with $\phi=14^{\circ}$.
It can be seen from the blue curve that although the beam covers several
points because of power leakage, we can locate the power peak of the
beam, which can be utilized for initial AoA estimation. The orange
curve demonstrates the effect of the optimal angle rotation for $\phi=14{}^{\circ}$.
It is obvious that more power is focused on $\phi=14{}^{\circ}$,
which makes the AoA estimation more accurate.

\begin{algorithm}
\caption{Common AoA Estimation}
\label{Algorithm-dft} \begin{algorithmic}[1] \REQUIRE $\mathbf{Y}_{1}$.

\STATE Calculate DFT: $\mathbf{Y}_{DFT}=\mathbf{U}_{N}^{\mathrm{H}}\mathbf{Y}_{1}$;

\STATE Calculate the power of each row: $\mathbf{z}(n)=||[\mathbf{Y}_{DFT}]_{n,:}||^{2},\forall n=1,2,\ldots,N$;

\STATE Find the rows with the power peak: $(\Omega_{N},\widehat{L})=\Gamma(\mathbf{z})$,
where $\Omega_{N}=\{n_{l},l=1,\cdots,\widehat{L}\}$;

\STATE Calculate the optimal angle rotation parameters $\{\triangle\widehat{\psi}_{l}\}_{l=1}^{\widehat{L}}$
via (\ref{eq:rotation});

\STATE Estimate AOAs for $1\leq l\leq\widehat{L}$: 
\begin{align}
 & \widehat{\phi}_{l}=\begin{cases}
\arccos\left(\frac{\lambda_{c}(n_{l}-1)}{d_{\mathrm{BS}}N}-\frac{\lambda_{c}\triangle\psi_{l}}{2\pi d_{\mathrm{BS}}}\right), & n_{l}\leq N\frac{d_{\mathrm{BS}}}{\lambda_{c}}\\
\arccos\left(\frac{\lambda_{c}(n_{l}-N-1)}{d_{\mathrm{BS}}N}-\frac{\lambda_{c}\triangle\psi_{l}}{2\pi d_{\mathrm{BS}}}\right), & n_{l}>N\frac{d_{\mathrm{BS}}}{\lambda_{c}}
\end{cases}.\label{eq:ini-aoa}
\end{align}

\ENSURE $\{\widehat{\phi}_{l}\}_{l=1}^{\widehat{L}}$. \end{algorithmic} 
\end{algorithm}

Algorithm \ref{Algorithm-dft} summarizes the estimation of the common
AoAs. After calculating the sum power of each row of $\mathbf{Y}_{DFT}$
in Step 2, we find the set of row indexes with peak power in Step
3. $\Gamma(\mathbf{z})$ denotes the operation of finding the indicies
with peak power in vector $\mathbf{z}$, $\Omega_{N}=\{n_{l},l=1,\cdots,\widehat{L}\}$
is a set to collect the indicies of the non-zero rows, and $\widehat{L}$
is the number of non-zero rows. We note that $\widehat{L}$ is the
estimated number of the propagation paths between the BS and the RIS,
and also the estimated number of common AoAs. For each $n_{l}$, Problem
(\ref{eq:rotation}) is solved to find the optimal angle rotation
parameter in Step 4. Finally, the common AOAs are estimated in Step
5.

\subsubsection{Estimation of the cascaded AoD cosines and gains}

With the estimated AoAs $\{\widehat{\phi}_{l}\}_{l=1}^{\widehat{L}}$
from Algorithm \ref{Algorithm-dft}, we obtain the estimated steering
matrix $\mathbf{\widehat{A}}_{N}=[\mathbf{a}_{N}(\widehat{\psi}_{1}),\ldots,\mathbf{a}_{N}(\widehat{\psi}_{\widehat{L}})]\in\mathbb{C}^{N\times\widehat{L}}$.
Based on the orthogonality of the massive steering matrix, i.e., $\mathbf{\widehat{A}}_{N}^{\mathrm{H}}\mathbf{A}_{N}\approx N\mathbf{I}_{L}$
due to Lemma \ref{othogonal}, the measurement matrix $\mathbf{Y}_{1}$
can be projected onto the common AoA steering matrix subspace as 
\begin{align}
\frac{1}{N\sqrt{p}}\mathbf{\widehat{A}}_{N}^{\mathrm{H}}\mathbf{Y}_{1} & \approx\boldsymbol{\Lambda}\mathbf{A}_{M}^{\mathrm{H}}\mathrm{Diag}\left(\mathbf{h}_{1}\right)\mathbf{E}_{1}+\frac{1}{N\sqrt{p}}\mathbf{\widehat{A}}_{N}^{\mathrm{H}}\mathbf{N}_{1}\nonumber \\
 & ={\bf H}_{\mathrm{RIS}}^{\mathrm{H}}\mathbf{E}_{1}+\frac{1}{N\sqrt{p}}\mathbf{\widehat{A}}_{N}^{\mathrm{H}}\mathbf{N}_{1},\label{eq:vv}
\end{align}
where ${\bf H}_{\mathrm{RIS}}=\mathrm{Diag}\left(\mathbf{h}_{1}^{*}\right)\mathbf{A}_{M}\boldsymbol{\Lambda}^{\mathrm{H}}$.
Based on (\ref{H2}) and (\ref{H3}), the $l$-th column of ${\bf H}_{\mathrm{RIS}}$
is given by 
\begin{equation}
{\bf h}_{\mathrm{RIS},l}=\mathrm{Diag}\{{\bf h}_{1}^{*}\}\mathbf{a}_{M}(\omega_{l})\alpha_{l}^{*},\label{hris}
\end{equation}
where ${\bf H}_{\mathrm{RIS}}=[\mathbf{h}_{\mathrm{RIS},1},\cdots,\mathbf{h}_{\mathrm{RIS},L}]$.
We claim that ${\bf h}_{\mathrm{RIS},l}$ can be estimated by transforming
each row of (\ref{eq:vv}) into a sparse signal recovery problem.
In particular, define $\frac{1}{N\sqrt{p}}\mathbf{\widehat{A}}_{N}^{\mathrm{H}}\mathbf{Y}_{1}=[\mathbf{p}_{1},\ldots,\mathbf{p}_{L}]^{\mathrm{H}}$,
where 
\begin{align}
\mathbf{p}_{l} & =\mathbf{E}_{1}^{\mathrm{H}}\mathbf{h}_{\mathrm{RIS},l}+\mathbf{n}_{\mathrm{noise}}\in\mathbb{C}^{\tau_{1}\times1}\nonumber \\
 & =\mathbf{E}_{1}^{\mathrm{H}}\mathrm{Diag}\{\mathbf{a}_{M}(\omega_{l})\}{\bf h}_{1}^{*}\alpha_{l}^{*}+\mathbf{n}_{\mathrm{noise}}\nonumber \\
 & =\mathbf{E}_{1}^{\mathrm{H}}\mathrm{Diag}\{\mathbf{a}_{M}(\omega_{l})\}\mathbf{A}_{M,1}^{*}\boldsymbol{\beta}_{1}^{*}\alpha_{l}^{*}+\mathbf{n}_{\mathrm{noise}}\nonumber \\
 & =\mathbf{E}_{1}^{\mathrm{H}}\left[\begin{array}{ccc}
\mathbf{a}_{M}\left(\omega_{l}-\varphi_{1,1}\right) & \cdots & \mathbf{a}_{M}\left(\omega_{l}-\varphi_{1,J_{1}}\right)\end{array}\right]\boldsymbol{\beta}_{1}^{*}\alpha_{l}^{*}+\mathbf{n}_{\mathrm{noise}}\label{eq:y1}
\end{align}
with $\mathbf{n}_{\mathrm{noise}}$ representing the corresponding
noise vector. To extract the cascaded directional cosine $\{\omega_{l}-\varphi_{1,j}\}_{j=1}^{J_{1}}$
and gains $\boldsymbol{\beta}_{1}^{*}\alpha_{l}^{*}$ from $\mathbf{p}_{l}$,
(\ref{eq:y1}) can be approximated by using the VAD representation
as 
\begin{align}
\mathbf{p}_{l} & =\mathbf{E}_{1}^{\mathrm{H}}\mathbf{A}\mathbf{b}_{l}+\mathbf{n}_{\mathrm{noise}},\label{pl}
\end{align}
where $\mathbf{A}\in\mathbb{C}^{M\times D}(M\ll D)$ is an overcomplete
dictionary matrix, each column of which represents the array steering
vector for possible values of $\omega_{l}-\varphi_{1,j}$. Since $\omega_{l}-\varphi_{1,j}\in[-2\frac{d_{\mathrm{RIS}}}{\lambda_{c}},2\frac{d_{\mathrm{RIS}}}{\lambda_{c}}]$,
$\mathbf{A}$ can be constructed as 
\begin{align}
\mathbf{A}=\biggl[  \mathbf{a}_{M}(-2\frac{d_{\mathrm{RIS}}}{\lambda_{c}}),\mathbf{a}_{M}((-2+\frac{4}{D})\frac{d_{\mathrm{RIS}}}{\lambda_{c}}),\ldots,
.  \mathbf{a}_{M}((2-\frac{4}{D})\frac{d_{\mathrm{RIS}}}{\lambda_{c}})\biggr].\label{diction}
\end{align}
Recall that $\boldsymbol{\beta}_{1}=[\beta_{1,1},\ldots,\beta_{1,J_{1}}]^{\mathrm{T}}$
in (\ref{h2}), $\mathbf{b}_{l}\in\mathbb{C}^{D\times1}$ is then
a sparse vector with $J_{1}$ cascaded gains $\{\alpha_{l}^{*}\beta_{1,j}^{*}\}_{j=1}^{J_{1}}$
as nonzero elements. Equation (\ref{pl}) can be cast as a sparse
signal recovery problem that can be solved using CS techniques, such
as OMP. Note that the phase shift matrix $\mathbf{E}_{1}$ in (\ref{pl})
will be designed for better estimation in Section \ref{training deng}.
It has been proved that $\tau_{1}\geq8J_{1}-2$ measurements are sufficient
to recover a $J_{1}$-sparse complex-valued signal vector \cite{pilot-overhead}.

However, if OMP is used $L$ times for solving $\mathbf{p}_{l}(1\leq l\leq L)$,
we need to estimate $J_{1}L$ independent sparse variables with high
complexity. In order to reduce the complexity, we exploit the following
scaling property. Specifically, we observe from (\ref{hris}) that
there is an angle and gain scaling between the cascaded multipaths
formed by different AoDs $\{\omega_{l}\}_{l=1}^{L}$ from the RIS.
That is, there is the following relationship between ${\bf h}_{\mathrm{RIS},l}$
and ${\bf h}_{\mathrm{RIS},r}$ for $1\leq l,r\leq L$: 
\begin{align}
{\bf h}_{\mathrm{RIS},l} & =\mathrm{Diag}\{\mathbf{a}_{M}(\omega_{l}-\omega_{r})\}\mathrm{Diag}\{{\bf h}_{1}^{*}\}\mathbf{a}_{M}(\omega_{r})\alpha_{r}^{*}\frac{\alpha_{l}^{*}}{\alpha_{r}^{*}}\nonumber \\
 & =\mathrm{Diag}\{\mathbf{a}_{M}(\omega_{l}-\omega_{r})\}{\bf h}_{\mathrm{RIS},r}\frac{\alpha_{l}^{*}}{\alpha_{r}^{*}}.\label{eq:proper}
\end{align}
Equation (\ref{eq:proper}) is called the angle-gain scaling property,
which implies that ${\bf h}_{\mathrm{RIS},l}$ for all $l$ can be
represented by one arbitrary ${\bf h}_{\mathrm{RIS},r}$. Let \begin{subequations}
\label{Problem-4} 
\begin{align}
\triangle\omega_{l} & =\omega_{l}-\omega_{r},\label{omiga}\\
x_{l} & =\frac{\alpha_{l}^{*}}{\alpha_{r}^{*}}.\label{alpha}
\end{align}
\end{subequations}Equation (\ref{eq:proper}) is then re-expressed
as ${\bf h}_{\mathrm{RIS},l}=\mathrm{Diag}\{{\bf h}_{\mathrm{RIS},r}\}\mathbf{a}_{M}(\triangle\omega_{l})x_{l}$.
Denote the estimate of ${\bf h}_{\mathrm{RIS},r}$ as $\widehat{{\bf h}}_{\mathrm{RIS},r}$
obtained from (\ref{pl}) using OMP. Further defining $\mathbf{z}_{l}(\triangle\omega_{l})=\mathbf{E}_{1}^{\mathrm{H}}\mathrm{Diag}\{\widehat{{\bf h}}_{\mathrm{RIS},r}\}\mathbf{a}_{M}(\triangle\omega_{l})$,
(\ref{eq:y1}) can be rewritten as 
\begin{equation}
\mathbf{p}_{l}=\mathbf{z}_{l}(\triangle\omega_{l})x_{l}+\mathbf{n}_{\mathrm{noise}}.\label{eq:y2}
\end{equation}

It is observed from (\ref{eq:y2}) that only two variables $\triangle\omega_{l}$
and $x_{l}$ need to be estimated. Since $\triangle\omega_{l}\in[-2\frac{d_{\mathrm{RIS}}}{\lambda_{c}},2\frac{d_{\mathrm{RIS}}}{\lambda_{c}}]$,
$\triangle\omega_{l}$ can then be estimated via a simple correlation-based
scheme 
\begin{equation}
\triangle\widehat{\omega}_{l}=\mathrm{arg}\max_{\triangle\omega\in[-2\frac{d_{\mathrm{RIS}}}{\lambda_{c}},2\frac{d_{\mathrm{RIS}}}{\lambda_{c}}]}\left|\left\langle \mathbf{p}_{l},\mathbf{z}_{l}(\triangle\omega)\right\rangle \right|.\label{eq:omiga}
\end{equation}
The parameter $x_{l}$ can be found as the solution of the LS problem
$\min_{x}||\mathbf{p}_{l}-\mathbf{z}_{l}(\triangle\widehat{\omega}_{l})x||_{2}$:
\begin{equation}
\widehat{x}_{l}=(\mathbf{z}_{l}^{\mathrm{H}}(\triangle\widehat{\omega}_{l})\mathbf{z}_{l}(\triangle\widehat{\omega}_{l}))^{-1}\mathbf{z}_{l}^{\mathrm{H}}(\triangle\widehat{\omega}_{l})\mathbf{p}_{l}.\label{eq:x}
\end{equation}

Let $\widehat{{\bf h}}_{\mathrm{RIS},l}=\mathrm{Diag}\{\widehat{{\bf h}}_{\mathrm{RIS},r}\}\mathbf{a}_{M}(\triangle\widehat{\omega}_{l})\widehat{x}_{l},(1\leq l\leq L,l\neq r)$,
so that the final estimated cascaded channel of user $1$ is given
by 
\begin{equation}
\widehat{\mathbf{G}}_{1}=\mathbf{\widehat{A}}_{N}\widehat{{\bf H}}_{\mathrm{R\mathrm{IS}}}^{\mathrm{H}},
\end{equation}
where $\widehat{{\bf H}}_{\mathrm{RIS}}=[\widehat{{\bf h}}_{\mathrm{RIS},1},\cdots,\widehat{{\bf h}}_{\mathrm{RIS},L}]$.

Algorithm \ref{Algorithm-G-1} summarizes the complete estimation
of $\mathbf{G}_{1}$. The common AoA steering matrix $\mathbf{A}_{N}$
is estimated by using the DFT and the angle rotation techniques in
Stage 1. In Stage 2 consisting of Steps 3-12, OMP is used to estimate
${\bf h}_{\mathrm{RIS},r}$. Here, $r$ is determined according to
Problem (\ref{eq:omiga-1}) such that the SNR of $\mathbf{p}_{r}$
is the maximum value among $\{\mathbf{p}_{l}\}_{l=1}^{L}$ (assuming
they have the same noise power) for better estimation accuracy for
the OMP method. The remaining ${\bf h}_{\mathrm{RIS},l}$ ($1\leq l\leq\widehat{L}$
and $l\neq r$) are estimated using the simple LS method and correlation-based
scheme in Stage 3 shown in Steps 13-16. Finally, we obtain the estimate
$\widehat{\mathbf{G}}_{1}=\mathbf{\widehat{A}}_{N}[\widehat{{\bf h}}_{\mathrm{RIS},1},\cdots,\widehat{{\bf h}}_{\mathrm{RIS},\widehat{L}}]^{\mathrm{H}}$.
The flow chart of Algorithm \ref{Algorithm-G-1} is shown in Fig.
\ref{stage}.

We emphasize that the cascaded AoD cosines and cascaded gains can
also been obtained in Algorithm \ref{Algorithm-G-1}, which facilitates
the cascaded channel estimation of other users in the next subsection.
In particular, the cascaded AoD cosines and cascaded gains from $\widehat{{\bf h}}_{\mathrm{RIS},r}$
in Step 12 are given by\begin{subequations} \label{Problem-5} 
\begin{align}
[\mathbf{a}_{M}(\widehat{\omega_{r}-\varphi_{1,1}})\cdots\mathbf{a}_{M}(\widehat{\omega_{r}-\varphi_{1,\widehat{J}_{1}}})] & =\mathbf{A}_{(:,\Omega_{i-1})},\label{eq:xs}\\
\widehat{\boldsymbol{\beta}_{1}^{*}\alpha_{r}^{*}} & =\mathbf{b}_{i-1}.\label{xs1}
\end{align}

\end{subequations} Based on (\ref{Problem-4}) and (\ref{Problem-5}),
the cascaded AoD cosines and cascaded gains from $\widehat{{\bf h}}_{\mathrm{RIS},l}$
($1\leq l\leq\widehat{L}$ and $l\neq r$) in Step 16 are given by\begin{subequations}
\label{Problem-6} 
\begin{align}
  [\mathbf{a}_{M}(\widehat{\omega_{l}-\varphi_{1,1}})\cdots\mathbf{a}_{M}(\widehat{\omega_{l}-\varphi_{1,\widehat{J}_{1}}})]&=\mathrm{Diag}\{\mathbf{a}_{M}(\triangle\widehat{\omega}_{l})\}\mathbf{A}_{(:,\Omega_{i-1})},\label{xs-1}\\
  \widehat{\boldsymbol{\beta}_{1}^{*}\alpha_{l}^{*}}&=\widehat{\boldsymbol{\beta}_{1}^{*}\alpha_{r}^{*}}\widehat{x}_{l}.\label{xs1-1}
\end{align}
\end{subequations}

Algorithm \ref{Algorithm-G-1} estimates $L$ AoAs in Stage 1, $J_{1}$
cascaded AoD cosines and $J_{1}$ cascaded gains in (\ref{Problem-5}),
and $2L-2$ scaling parameters in Step 14 and Step 15. Therefore,
Algorithm \ref{Algorithm-G-1} uses a total of $\tau_{1}\geq8J_{1}-2$
time slots to estimate $3L+2J_{1}-2$ parameters to recover channel
$\mathbf{G}_{1}$ of dimension $N\times M$. Note that the number
of time slots required is not related to $L$, which evidences the
advantage of our proposed estimation method.

\begin{algorithm}
\caption{DFT-OMP-based Estimation of $\mathbf{G}_{1}$}
\label{Algorithm-G-1} \begin{algorithmic}[1] \REQUIRE $\mathbf{Y}_{1}$,
$\mathbf{A}$.

\STATE \textbf{Stage 1: }Return estimated common AoA steering matrix
$\mathbf{\widehat{A}}_{N}$ and $\widehat{L}$ using Algorithm \ref{Algorithm-dft}.

\STATE Calculate $[\mathbf{p}_{1},\ldots,\mathbf{p}_{\widehat{L}}]=\frac{1}{N\sqrt{p}}\mathbf{Y}_{1}^{\mathrm{H}}\mathbf{\widehat{A}}_{N}$.

\STATE \textbf{Stage 2: }Estimate ${\bf h}_{\mathrm{RIS},r}$ from
$\mathbf{p}_{r}$ using the OMP algorithm, where $r$ is determined
according to 
\begin{equation}
r=\mathrm{arg}\max_{1\leq i\leq\widehat{L}}||\mathbf{p}_{i}||^{2}.\label{eq:omiga-1}
\end{equation}

\STATE Calculate equivalent dictionary $\mathbf{D}=\mathbf{E}_{1}^{\mathrm{H}}\mathbf{A}$.

\STATE Initialize $\Omega_{0}=\emptyset$, $\mathbf{r}_{0}=\mathbf{p}_{r}$,
$i=1$.

\REPEAT

\STATE $d_{i}=\mathrm{arg}\max_{d=1,2,\ldots,D}|\mathbf{D}_{(:,d)}^{\mathrm{H}}\mathbf{r}_{i-1}|.$

\STATE $\Omega_{i}=\Omega_{i-1}\cup d_{i}.$

\STATE LS solution: $\mathbf{b}_{i}=(\mathbf{D}_{(:,\Omega_{i})}^{\mathrm{H}}\mathbf{D}_{(:,\Omega_{i})})^{-1}\mathbf{D}_{(:,\Omega_{i})}^{\mathrm{H}}\mathbf{p}_{r}.$

\STATE $\mathbf{r}_{i}=\mathbf{p}_{r}-\mathbf{D}_{(:,\Omega_{i})}\mathbf{b}_{i}$.

\STATE $i=i+1$.

\UNTIL $||\mathbf{r}_{i-1}||_{2}\leq$threshold.

\STATE Obtain the estimates: \begin{subequations} \label{Problem-1}
\begin{align}
 & \widehat{J}_{1}=i-1,\\
 & \widehat{{\bf h}}_{\mathrm{RIS},r}=\mathbf{A}_{(:,\Omega_{i-1})}\mathbf{b}_{i-1}.\label{xs3}
\end{align}
\end{subequations}

\STATE \textbf{Stage 3: }Estimate ${\bf h}_{\mathrm{RIS},l}$ from
$\mathbf{p}_{l}$ for $1\leq l\leq\widehat{L}$ and $l\neq r$:

\STATE Calculate $\triangle\widehat{\omega}_{l}$ according to (\ref{eq:omiga}).

\STATE Calculate $\widehat{x}_{l}$ according to (\ref{eq:x}).

\STATE Obtain the estimates for $1\leq l\leq\widehat{L}$ and $l\neq r$:
\begin{align}
 & \widehat{{\bf h}}_{\mathrm{RIS},l}=\mathrm{Diag}\{\widehat{{\bf h}}_{\mathrm{RIS},r}\}\mathbf{a}_{M}(\triangle\widehat{\omega}_{l})\widehat{x}_{l}.\label{xs3-1}
\end{align}

\ENSURE $\widehat{\mathbf{G}}_{1}=\mathbf{\widehat{A}}_{N}[\widehat{{\bf h}}_{\mathrm{RIS},1},\cdots,\widehat{{\bf h}}_{\mathrm{RIS},\widehat{L}}]^{\mathrm{H}}$.
\end{algorithmic} 
\end{algorithm}

\subsection{Channel Estimation for Other Users in the First Coherence Block}

Algorithm \ref{Algorithm-G-1} can also be used for the channel estimation
of the other users, where Stage 1 can be omitted because all users
share the common AoA steering matrix $\mathbf{\widehat{A}}_{N}$.
In addition to $\mathbf{\widehat{A}}_{N}$, all users also share the
common matrices $\boldsymbol{\Lambda}$ and $\mathbf{A}_{M}$ in their
channel matrices $\mathbf{G}_{k},\forall k$. Note that for the cascaded
channel $\mathbf{G}_{k}=\mathbf{H}\mathrm{Diag}(\mathbf{h}_{k}),2\leq k\leq K$
in (\ref{eq:3}), we might expect that if the common channel $\mathbf{H}$
is known, channel $\mathbf{h}_{k}$ can be readily estimated using
a sparse signal recovery problem. However, it is intractable to obtain
$\mathbf{H}$ from the estimated cascaded channel $\widehat{\mathbf{G}}_{1}$
due to the coupling of angles $\cos(\theta_{l})-\cos(\vartheta_{1,j})$
and channel gains $\alpha_{l}\beta_{1,j}$ with each cascaded subpath
of user $1$. However, we can construct a substitute for $\mathbf{H}$
(denoted by $\mathbf{H}_{\mathrm{c}}$) by only using $\widehat{\mathbf{G}}_{1}$.
The substitute $\mathbf{H}_{\mathrm{c}}$ contains reparameterized
information about $\mathbf{H}$. Then, (\ref{eq:3}) can be rewritten
as 
\begin{equation}
\mathbf{G}_{k}=\mathbf{H}_{\mathrm{c}}\mathrm{Diag}(\mathbf{h}_{\mathrm{c},k}),2\leq k\leq K,\label{paril-G}
\end{equation}
where $\mathbf{h}_{\mathrm{c},k}$ is the corresponding reparameterized
CSI of $\mathbf{h}_{k}$. In the following, we first construct $\mathbf{H}_{\mathrm{c}}$
based on the estimated channel information from Algorithm 2 and then
estimate the reparameterized channel information $\mathbf{h}_{\mathrm{c},k}$.

\subsubsection{Construction of $\mathbf{H}_{\mathrm{c}}$}

In the following, we show how to construct $\mathbf{H}_{\mathrm{c}}$
by exploiting the structure of $\widehat{\mathbf{G}}_{1}$. In particular,
(\ref{eq:H1-1}) is reformulated as 
\begin{align}
\mathbf{H} & =\mathbf{A}_{N}\boldsymbol{\Lambda}\mathbf{A}_{M}^{\mathrm{H}}\nonumber \\
 & =\mathbf{A}_{N}\frac{1}{\overline{\beta}}\boldsymbol{\Lambda}_{\mathrm{c}}\mathbf{A}_{\mathrm{c}}^{\mathrm{H}}\mathrm{Diag}(\mathbf{a}_{M}(\overline{\varphi}))\nonumber \\
 & =\frac{1}{\overline{\beta}}\mathbf{H}_{\mathrm{c}}\mathrm{Diag}\left(\mathbf{a}_{M}(\overline{\varphi})\right),\label{H3-1}
\end{align}
with \begin{subequations} \label{hc} 
\begin{align}
\mathbf{H}_{\mathrm{c}} & =\mathbf{A}_{N}\boldsymbol{\Lambda}_{\mathrm{c}}\mathbf{A}_{\mathrm{c}}^{\mathrm{H}},\label{b2}\\
\boldsymbol{\Lambda}_{\mathrm{c}} & =\overline{\beta}\boldsymbol{\Lambda},\label{b3}\\
\mathbf{A}_{\mathrm{c}} & =\mathrm{Diag}(\mathbf{a}_{M}(\overline{\varphi}))\mathbf{A}_{M},\label{b4}\\
\overline{\varphi} & =-\frac{1}{J_{1}}\sum_{j=1}^{J_{1}}\varphi_{1,j},\label{b5}\\
\overline{\beta} & =\frac{1}{J_{1}}\mathbf{1}_{J_{1}}^{\mathrm{T}}\boldsymbol{\beta}_{1},\label{b6}
\end{align}
\end{subequations}where $\mathbf{1}_{J_{1}}$ is an all-one vector
with dimension $J_{1}\times1$, and $\boldsymbol{\beta}_{1}$ is defined
in (\ref{h2}).

Using (\ref{H2}), (\ref{alpha}) and (\ref{b6}), $\boldsymbol{\Lambda}_{\mathrm{c}}$
in (\ref{b3}) can be re-expressed as\begin{subequations} \label{hc-x-2}
\begin{align}
\boldsymbol{\Lambda}_{\mathrm{c}} & =\overline{\beta}\boldsymbol{\Lambda}\nonumber \\
 & =\overline{\beta}\mathrm{Diag}(\alpha_{1},\ldots,\alpha_{L})\nonumber \\
 & =\left[\overline{\beta}^{*}\alpha_{r}^{*}\mathrm{Diag}(x_{1},\ldots,x_{L})\right]^{*}\label{er1}\\
 & =\left[\frac{1}{J_{1}}\mathbf{1}_{J_{1}}^{\mathrm{T}}\boldsymbol{\beta}_{1}^{*}\alpha_{r}^{*}\mathrm{Diag}(x_{1},\ldots,x_{L})\right]^{*},\label{r32}
\end{align}
\end{subequations}where the estimate of $\boldsymbol{\beta}_{1}^{*}\alpha_{r}^{*}$
is given in (\ref{xs1}), and the estimate of $[x_{1},x_{2},\ldots,x_{L}]$
is given in (\ref{eq:x}). Then, the estimate of $\boldsymbol{\Lambda}_{\mathrm{c}}$
is obtained as 
\begin{equation}
\widehat{\boldsymbol{\Lambda}}_{\mathrm{c}}=\left[\frac{1}{J_{1}}\mathbf{1}_{J_{1}}^{\mathrm{T}}\widehat{\boldsymbol{\beta}_{1}^{*}\alpha_{r}^{*}}\mathrm{Diag}([\widehat{x}_{1},\ldots,\widehat{x}_{\widehat{L}}])\right]^{*}.\label{eq:ac}
\end{equation}
For $\mathbf{A}_{\mathrm{c}}$, by substituting (\ref{H3}), (\ref{omiga})
and (\ref{b5}) into (\ref{b4}), we have\begin{subequations} \label{hc-a}
\begin{align*}
 \mathbf{A}_{\mathrm{c}}
 & =\mathrm{Diag}(\mathbf{a}_{M}(\overline{\varphi}))\mathbf{A}_{M}\\
 & =\mathrm{Diag}(\mathbf{a}_{M}(\overline{\varphi}))[\mathbf{a}_{M}(\omega_{1}),\ldots,\mathbf{a}_{M}(\omega_{L})]\\
 & =\mathrm{Diag}(\mathbf{a}_{M}(\omega_{r}+\overline{\varphi}))[\mathbf{a}_{M}(\triangle\omega_{1}),\ldots,\mathbf{a}_{M}(\triangle\omega_{L})]\\
 & =\mathrm{Diag}(\mathbf{a}_{M}(\omega_{r}-\frac{1}{J_{1}}\sum_{j=1}^{J_{1}}\varphi_{1,j}))[\mathbf{a}_{M}(\triangle\omega_{1}),\ldots,\mathbf{a}_{M}(\triangle\omega_{L})]\\
 & =\mathrm{Diag}(\mathbf{a}_{M}(\frac{1}{J_{1}}\sum_{j=1}^{J_{1}}(\omega_{r}-\varphi_{1,j})))[\mathbf{a}_{M}(\triangle\omega_{1}),\ldots,\mathbf{a}_{M}(\triangle\omega_{L})],
\end{align*}
\end{subequations}where the estimate of $\{\omega_{r}-\varphi_{1,j}\}_{j=1}^{J_{1}}$
and $\{\triangle\omega_{l}\}_{l=1}^{L}$ are given by (\ref{eq:omiga})
and (\ref{eq:xs}), respectively. Then, we can obtain the estimate
of $\mathbf{A}_{\mathrm{c}}$ as 
\begin{align}
\widehat{\mathbf{A}}_{\mathrm{c}}= & \mathrm{Diag}(\mathbf{a}_{M}(\frac{1}{\widehat{J}_{1}}\sum_{j=1}^{\widehat{J}_{1}}(\widehat{\omega_{r}-\varphi_{1,j}}))) [\mathbf{a}_{M}((\triangle\widehat{\omega}_{1}),\cdots\mathbf{a}_{M}(\triangle\widehat{\omega}_{\widehat{L}})]\nonumber \\
= & \mathrm{Diag}(\mathbf{a}_{M}(\widehat{\omega_{r}+\overline{\varphi}}))[\mathbf{a}_{M}((\triangle\widehat{\omega}_{1}),\cdots\mathbf{a}_{M}(\triangle\widehat{\omega}_{\widehat{L}})].\label{eq:av}
\end{align}
With $\widehat{\mathbf{A}}_{N}$, (\ref{eq:ac}) and (\ref{eq:av}),
the estimate of $\mathbf{H}_{\mathrm{c}}$ is given by $\widehat{\mathbf{H}}_{\mathrm{c}}=\widehat{\mathbf{A}}_{N}\widehat{\boldsymbol{\Lambda}}_{\mathrm{c}}\widehat{\mathbf{A}}_{\mathrm{c}}^{\mathrm{H}}$.

\subsubsection{Estimation of reparameterized CSI $\mathbf{h}_{\mathrm{c},k}$}

In this subsection, we discuss how to use the reparameterized common
channel $\mathbf{H}_{\mathrm{c}}$ to help the channel estimation
of other users with low pilot overhead. In particular, by substituting
$\mathbf{H}=\frac{1}{\overline{\beta}}\mathbf{H}_{\mathrm{c}}\mathrm{Diag}(\mathbf{a}_{M}(\overline{\varphi}))$
in (\ref{H3-1}) into (\ref{eq:3}), we have 
\begin{align}
\mathbf{G}_{k} & =\mathbf{H}\mathrm{Diag}(\mathbf{h}_{k})\nonumber \\
 & =\frac{1}{\overline{\beta}}\mathbf{H}_{\mathrm{c}}\mathrm{Diag}\left(\mathbf{a}_{M}(\overline{\varphi})\right)\mathrm{Diag}(\mathbf{h}_{k})\nonumber \\
 & =\frac{1}{\overline{\beta}}\mathbf{H}_{\mathrm{c}}\mathrm{Diag}\left(\mathrm{Diag}\left(\mathbf{a}_{M}(\overline{\varphi})\right)\mathbf{h}_{k}\right)\nonumber \\
 & =\mathbf{H}_{\mathrm{c}}\mathrm{Diag}(\mathbf{h}_{\mathrm{c},k}),\label{Gc}
\end{align}
where 
\begin{equation}
\mathbf{h}_{\mathrm{c},k}=\frac{1}{\overline{\beta}}\mathrm{Diag}\left(\mathbf{a}_{M}(\overline{\varphi})\right)\mathbf{h}_{k}\label{fe-1}
\end{equation}
contains reparameterized CSI of $\mathbf{h}_{k}$.

Similar to (\ref{eq:vv}), $\mathbf{Y}_{k}$ in (\ref{eq:m-y-2})
is first projected onto the common AoA steering matrix subspace as
\begin{align}
\frac{1}{N\sqrt{p}}\mathbf{\widehat{A}}_{N}^{\mathrm{H}}\mathbf{Y}_{k} & =\frac{1}{N\sqrt{p}}\mathbf{\widehat{A}}_{N}^{\mathrm{H}}(\sqrt{p}\mathbf{H}_{\mathrm{c}}\mathrm{Diag}(\mathbf{h}_{\mathrm{c},k})\mathbf{E}_{k}+\mathbf{N}_{k})\nonumber \\
 & \approx\boldsymbol{\Lambda}_{\mathrm{c}}\mathbf{A}_{\mathrm{c}}^{\mathrm{H}}\mathrm{Diag}(\mathbf{h}_{\mathrm{c},k})\mathbf{E}_{k}+\frac{1}{N\sqrt{p}}\mathbf{\widehat{A}}_{N}^{\mathrm{H}}\mathbf{N}_{k}.\label{mm}
\end{align}
Recall that $\mathbf{E}_{k}=[\mathbf{e}_{1},\ldots,\mathbf{e}_{\tau_{k}}]$
in (\ref{cd}). By vectorizing (\ref{mm}) and defining $\mathbf{z}_{k}=\mathrm{vec}(\frac{1}{N\sqrt{p}}\mathbf{\widehat{A}}_{N}^{\mathrm{H}}\mathbf{Y}_{k})\in\mathbb{C}^{\tau_{k}L\times1}$,
we have 
\begin{align}
\mathbf{z}_{k} & =\left[\begin{array}{c}
\boldsymbol{\Lambda}_{\mathrm{c}}\mathbf{A}_{\mathrm{c}}^{\mathrm{H}}\mathrm{Diag}(\mathbf{h}_{\mathrm{c},k})\mathbf{e}_{1}\\
\vdots\\
\boldsymbol{\Lambda}_{\mathrm{c}}\mathbf{A}_{\mathrm{c}}^{\mathrm{H}}\mathrm{Diag}(\mathbf{h}_{\mathrm{c},k})\mathbf{e}_{\tau_{k}}
\end{array}\right]+\mathbf{n}_{\mathrm{noise}}\nonumber \\
 & =\mathbf{Z}_{k}\mathbf{h}_{\mathrm{c},k}+\mathbf{n}_{\mathrm{noise}},\label{zz}
\end{align}
where $\mathbf{n}_{\mathrm{noise}}$ represents the corresponding
noise and 
\begin{equation}
\mathbf{Z}_{k}=\left[\begin{array}{c}
\boldsymbol{\Lambda}_{\mathrm{c}}\mathbf{A}_{\mathrm{c}}^{\mathrm{H}}\mathrm{Diag}\left(\mathbf{e}_{1}\right)\\
\vdots\\
\boldsymbol{\Lambda}_{\mathrm{c}}\mathbf{A}_{\mathrm{c}}^{\mathrm{H}}\mathrm{Diag}\left(\mathbf{e}_{\tau_{k}}\right)
\end{array}\right].\label{eq:zz}
\end{equation}

By replacing $\mathbf{h}_{k}$ with $\mathbf{h}_{k}=\mathbf{A}_{M,k}\boldsymbol{\beta}_{k}$
from (\ref{eq:H-1-1}), $\mathbf{h}_{\mathrm{c},k}$ in (\ref{fe-1})
can be unfolded as 
\begin{align}
\mathbf{h}_{\mathrm{c},k} & =\frac{1}{\overline{\beta}}\mathrm{Diag}\left(\mathbf{a}_{M}(\overline{\varphi})\right)\mathbf{h}_{k}\nonumber \\
 & =\frac{1}{\overline{\beta}}[\mathbf{a}_{M}\left(\varphi_{k,1}+\overline{\varphi}\right)\cdots\mathbf{a}_{M}\left(\varphi_{k,J_{k}}+\overline{\varphi}\right)]\boldsymbol{\beta}_{k}.\label{co}
\end{align}
Since $\varphi_{k,1}+\overline{\varphi}\in[-2\frac{d_{\mathrm{RIS}}}{\lambda_{c}},2\frac{d_{\mathrm{RIS}}}{\lambda_{c}}]$,
(\ref{co}) can be further approximated by using the VAD representation
as 
\begin{align}
\mathbf{h}_{\mathrm{c},k} & =\mathbf{A}\mathbf{c}_{k},\label{ko}
\end{align}
where $\mathbf{A}$ is defined in (\ref{diction}), and $\mathbf{c}_{k}\in\mathbb{C}^{G\times1}$
is a sparse vector with $J_{k}$ gains $\{\frac{1}{\overline{\beta}}\beta_{k,j}\}_{j=1}^{J_{k}}$
as the nonzero elements.

With (\ref{ko}), (\ref{zz}) can be approximated as a sparse signal
recovery problem 
\begin{equation}
\mathbf{z}_{k}=\mathbf{Z}_{k}\mathbf{A}\mathbf{c}_{k}+\mathbf{n}_{\mathrm{noise}}.\label{eq:zc}
\end{equation}
Note that $\mathbf{Z}_{k}$ is determined using (\ref{eq:ac}) and
(\ref{eq:av}). Hence, Problem (\ref{eq:zc}) could be solved by using
CS technique, such as OMP. Note that the phase shift vectors $\{\mathbf{e}_{t}\}_{t=1}^{\tau_{k}}$
in $\mathbf{Z}_{k}$ will be designed in Section \ref{training deng}
to achieve high estimation accuracy.

Algorithm \ref{Algorithm-G-2} summarizes the OMP-based estimation
of $\mathbf{G}_{k},2\leq k\leq K$. To effectively recover the $J_{k}$-sparse
signal $\mathbf{c}_{k}$, the dimension of $\mathbf{z}_{k}\in\mathbb{C}^{\tau_{k}L\times1}$
should satisfy the requirement $\tau_{k}L\geq8J_{k}-2$ \cite{pilot-overhead}.
Thus, the pilot overhead required by user $k$ is $\tau_{k}\geq(8J_{k}-2)/L$.

\begin{algorithm}
\caption{Estimation of $\mathbf{G}_{k},2\protect\leq k\protect\leq K$}
\label{Algorithm-G-2} \begin{algorithmic}[1] \REQUIRE $\mathbf{A}$,
$\mathbf{Y}_{k},2\leq k\leq K$.

\STATE Return $\mathbf{\widehat{A}}_{N}$ from Algorithm \ref{Algorithm-dft}.

\STATE Construct $\widehat{\boldsymbol{\Lambda}}_{\mathrm{c}}$ according
to (\ref{eq:ac}).

\STATE Construct $\widehat{\mathbf{A}}_{\mathrm{c}}$ according to
(\ref{eq:av}).

\STATE Calculate equivalent dictionary $\mathbf{R}=\mathbf{Z}_{k}\mathbf{A}$
according to (\ref{eq:zz}) with $\widehat{\boldsymbol{\Lambda}}_{\mathrm{c}}$
and $\widehat{\mathbf{A}}_{\mathrm{c}}$.

\FOR{$2\leq k\leq K$}

\STATE Calculate $\mathbf{z}_{k}=\mathrm{vec}(\frac{1}{N\sqrt{p}}\mathbf{\widehat{A}}_{N}^{\mathrm{H}}\mathbf{Y}_{k})$.

\STATE Initialize $\Omega_{0}=\emptyset$, $\mathbf{r}_{0}=\mathbf{z}_{k}$,
$i=1$.

\REPEAT

\STATE $d_{i}=\mathrm{arg}\max_{d=1,2,\ldots,D}|\mathbf{R}_{(:,d)}^{\mathrm{H}}\mathbf{r}_{i-1}|.$

\STATE $\Omega_{i}=\Omega_{i-1}\cup d_{i}.$

\STATE LS solution: $\mathbf{b}_{i}=(\mathbf{R}_{(:,\Omega_{i})}^{\mathrm{H}}\mathbf{R}_{(:,\Omega_{i})})^{-1}\mathbf{R}_{(:,\Omega_{i})}^{\mathrm{H}}\mathbf{p}_{r}.$

\STATE $\mathbf{r}_{i}=\mathbf{p}_{r}-\mathbf{R}_{(:,\Omega_{i})}\mathbf{b}_{i}$.

\STATE $i=i+1$.

\UNTIL $||\mathbf{r}_{i-1}||_{2}\leq$threshold.

\STATE Calculate the estimated reparameterized common channel $\widehat{\mathbf{H}}_{\mathrm{c}}=\widehat{\mathbf{A}}_{N}\widehat{\boldsymbol{\Lambda}}_{\mathrm{c}}\widehat{\mathbf{A}}_{\mathrm{c}}^{\mathrm{H}}$.

\STATE Obtain the estimates: \begin{subequations} \label{Problem-1-1}
\begin{align}
 & \widehat{\mathbf{h}}_{\mathrm{c},k}=\mathbf{A}_{(:,\Omega_{i-1})}\mathbf{b}_{i-1},\label{xs3-2}\\
 & \widehat{\mathbf{G}}_{k}=\widehat{\mathbf{H}}_{\mathrm{c}}\mathrm{Diag}(\widehat{\mathbf{h}}_{\mathrm{c},k}).
\end{align}
\end{subequations}

\ENDFOR

\ENSURE $\widehat{\mathbf{G}}_{k},2\leq k\leq K$. \end{algorithmic} 
\end{algorithm}

We highlight the fact that the cascaded AoD cosines can also be obtained
after $[\mathbf{a}_{M}(\widehat{\varphi_{k,1}+\overline{\varphi}})\cdots \\
\mathbf{a}_{M}(\widehat{\varphi_{k,J_{k}}+\overline{\varphi}})]=\mathbf{A}_{(:,\Omega_{i-1})}$
is determined from (\ref{co}) when using OMP, which facilitates the
cascaded channel estimation in the subsequent channel coherence blocks
in the next subsection. In particular, the cascaded AoD cosines of
user $k$ for $2\leq k\leq K$ and $1\leq l\leq L$ are given by 
\begin{align}
 & [\mathbf{a}_{M}(\widehat{\omega_{l}-\varphi_{k,1}})\cdots\mathbf{a}_{M}(\widehat{\omega_{l}-\varphi_{k,J_{k}}})]\nonumber \\
 & =\mathrm{Diag}(\mathbf{a}_{M}(\widehat{\omega_{l}+\overline{\varphi}}))[\mathbf{a}_{M}^{*}(\widehat{\varphi_{k,1}+\overline{\varphi}})\cdots\mathbf{a}_{M}^{*}(\widehat{\varphi_{k,J_{k}}+\overline{\varphi}})]\nonumber \\
 & =\mathrm{Diag}(\widehat{\mathbf{A}}_{\mathrm{c}(:,l)})[\mathbf{a}_{M}^{*}(\widehat{\varphi_{k,1}+\overline{\varphi}})\cdots\mathbf{a}_{M}^{*}(\widehat{\varphi_{k,J_{k}}+\overline{\varphi}})],\label{BB}
\end{align}
where $\widehat{\mathbf{A}}_{\mathrm{c}(:,l)}$ is given in (\ref{eq:av}).

\subsection{Channel Estimation in the Remaining Coherence Blocks}

The channel gains need to be re-estimated for the remaining channel
coherence blocks as shown in Fig. \ref{channel estimation}. With
knowledge of the angle information obtained in the first coherence
block, only the cascaded channel gains need to be re-estimated.

For the remaining coherence blocks, the measurement matrix for user
$k$ at the BS in (\ref{eq:m-y-2}) is considered again: 
\begin{equation}
\mathbf{Y}_{k}=\sqrt{p}\mathbf{G}_{k}\mathbf{E}_{k}+\mathbf{N}_{k}\in\mathbb{C}^{N\times\tau_{k}}.\label{eq:m-y-2-1}
\end{equation}
Following the same derivations as in (\ref{eq:vv}) and (\ref{eq:y1}),
we define $\frac{1}{N\sqrt{p}}\mathbf{\widehat{A}}_{N}^{\mathrm{H}}\mathbf{Y}_{k}=[\mathbf{q}_{k,1},\ldots,\mathbf{q}_{k,L}]^{\mathrm{H}}$,
where 
\begin{align}
\mathbf{q}_{k,l} & =\mathbf{E}_{k}^{\mathrm{H}}\mathbf{B}_{k,l}\boldsymbol{\beta}_{k}^{*}\alpha_{l}^{*}+\mathbf{n}_{\mathrm{noise}},\label{eq:Y1-1}
\end{align}
$\mathbf{B}_{k,l}=[\mathbf{a}_{M}(\omega_{l}-\varphi_{k,1})\cdots\mathbf{a}_{M}(\omega_{l}-\varphi_{k,J_{k}})]$,
and $\mathbf{n}_{\mathrm{noise}}$ represents the corresponding noise
vector. Denote the estimate of $\mathbf{B}_{k,l}$ as $\widehat{\mathbf{B}}_{k,l}=[\mathbf{a}_{M}(\widehat{\omega_{l}-\varphi_{k,1}})\cdots\mathbf{a}_{M}(\widehat{\omega_{l}-\varphi_{k,J_{k}}})]$
obtained from (\ref{eq:xs}), (\ref{xs-1}) for $k=1$, and from (\ref{BB})
for $2\leq k\leq K$ in the first coherence block. Then the LS estimate
of $\boldsymbol{\beta}_{k}^{*}\alpha_{l}^{*}$ is given by 
\begin{align}
\widehat{\boldsymbol{\beta}_{k}^{*}\alpha_{l}^{*}} & =(\widehat{\mathbf{B}}_{k,l}^{\mathrm{H}}\mathbf{E}_{k}\mathbf{E}_{k}^{\mathrm{H}}\widehat{\mathbf{B}}_{k,l})^{-1}\widehat{\mathbf{B}}_{k,l}^{\mathrm{H}}\mathbf{E}_{k}\mathbf{q}_{k,l},\label{eq:1a}
\end{align}
Note that $\mathbf{E}_{k}^{\mathrm{H}}\widehat{\mathbf{B}}_{k,l}\in\mathbb{C}^{\tau_{k}\times J_{k}}$
must be a matrix with full row rank to ensure the feasibility of the
pseudo inverse operation in (\ref{eq:1a}), which means the pilot
length must satisfy $\tau_{k}\geq J_{k}$ for user $k$.

Define $\widehat{{\bf H}}_{\mathrm{RIS},k}=[\widehat{\mathbf{B}}_{k,1}\widehat{\boldsymbol{\beta}_{k}^{*}\alpha_{1}^{*}},\ldots,\widehat{\mathbf{B}}_{k,L}\widehat{\boldsymbol{\beta}_{k}^{*}\alpha_{L}^{*}}]$.
The uplink channel of the $k$-th user can then be reconstructed using
the updated cascaded channel gains obtained in this coherence block
and the angle information obtained during the first coherence block
as 
\begin{align}
\widehat{\mathbf{G}}_{k}=\mathbf{\widehat{A}}_{N}\widehat{{\bf H}}_{\mathrm{RIS},k}^{\mathrm{H}}.\label{GK}
\end{align}

\section{Training reflection coefficient optimization\label{training deng}}

The performance of OMP-based channel estimation is positively related
to the orthogonality of its equivalent dictionary. Therefore, in this
section, we optimize the training phase shift matrices to generate
approximately orthogonal equivalent dictionaries. Specifically, $\mathbf{E}_{k},\forall k\in\mathcal{K}$
are designed to improve the ability of OMP to recover the sparsest
signals $\mathbf{b}_{l}$ and $\mathbf{c}_{k}$ from the sparse recovery
problems $\mathbf{p}_{l}=\mathbf{E}_{1}^{\mathrm{H}}\mathbf{A}\mathbf{b}_{l}+\mathbf{n}_{\mathrm{noise}}$
in (\ref{pl}) and $\mathbf{z}_{k}=\mathbf{Z}_{k}\mathbf{A}\mathbf{c}_{k}+\mathbf{n}_{\mathrm{noise}}$
in (\ref{eq:zc}), respectively. In the following, we first investigate
the design of $\mathbf{E}_{1}$ in (\ref{pl}), and then extend the
solution to the design of $\mathbf{E}_{k}$ $(2\leq k\leq K)$.

Our approach is motivated by the theoretical work of \cite{OMP-TIT}
which shows that the sparse signal $\mathbf{b}_{l}$ can be recovered
successfully by OMP only when the following condition holds: 
\begin{equation}
||\mathbf{b}_{l}||_{0}\leq\frac{1}{2}\left(1+\frac{1}{\mu}\right),\label{eq:b1}
\end{equation}
where $\mu$ is the mutual coherence of the equivalent dictionary
$\mathbf{D}=\mathbf{E}_{1}^{\mathrm{H}}\mathbf{A}$ defined by 
\begin{equation}
\mu=\max_{i\neq j}\frac{|\mathbf{D}_{(:,i)}^{\mathrm{H}}\mathbf{D}_{(:,j)}|}{||\mathbf{D}_{(:,i)}||_{2}||\mathbf{D}_{(:,j)}||_{2}}.
\end{equation}
The condition in (\ref{eq:b1}) suggests that $\mathbf{D}$ should
be as incoherent (orthogonal) as possible, which leads to the following
design problem 
\begin{align}
\min_{\mathbf{E}_{1}} & ||\mathbf{D}^{\mathrm{H}}\mathbf{D}-\mathbf{I}_{D}||_{F}^{2}\nonumber \\
\mathrm{s.t.} & |[\mathbf{E}_{1}]_{m,n}|=1,1\leq m\leq M,1\leq n\leq\tau_{1}.\label{eq:kk}
\end{align}

The solution for the unconstrained version of Problem (\ref{eq:kk})
has been investigated in \cite{Sensing-TSP}, and the method designed
therein is extended to solve the constrained Problem (\ref{eq:kk})
in \cite{ris-omp-2}. Based on \cite{ris-omp-2} and \cite{Sensing-TSP},
we propose a more concise solution in the following. To begin, note
that 
\begin{align}
 & ||\mathbf{D}^{\mathrm{H}}\mathbf{D}-\mathbf{I}_{D}||_{F}^{2}\nonumber \\
= & \mathrm{tr}\{\mathbf{D}^{\mathrm{H}}\mathbf{D}\mathbf{D}^{\mathrm{H}}\mathbf{D}-2\mathbf{D}^{\mathrm{H}}\mathbf{D}+\mathbf{I}_{D}\}\nonumber \\
= & \mathrm{tr}\{\mathbf{D}\mathbf{D}^{\mathrm{H}}\mathbf{D}\mathbf{D}^{\mathrm{H}}-2\mathbf{D}\mathbf{D}^{\mathrm{H}}+\mathbf{I}_{\tau_{1}}\}+(D-\tau_{1})\nonumber \\
= & ||\mathbf{D}\mathbf{D}^{\mathrm{H}}-\mathbf{I}_{\tau_{1}}||_{F}^{2}+(D-\tau_{1}).\label{d3}
\end{align}
Using (\ref{d3}), Problem (\ref{eq:kk}) reduces to 
\begin{align}
\min_{\mathbf{E}_{1}} & ||\mathbf{D}\mathbf{D}^{\mathrm{H}}-\mathbf{I}_{\tau_{1}}||_{F}^{2}=||\mathbf{E}_{1}^{\mathrm{H}}\mathbf{A}\mathbf{A}^{\mathrm{H}}\mathbf{E}_{1}-\mathbf{I}_{\tau_{1}}||_{F}^{2}\nonumber \\
\mathrm{s.t.} & |[\mathbf{E}_{1}]_{m,n}|=1,1\leq m\leq M,1\leq n\leq\tau_{1}.\label{eq:kk-1}
\end{align}
Define the eigenvalue decomposition $\mathbf{A}\mathbf{A}^{\mathrm{H}}=\mathbf{U}\boldsymbol{\Upsilon}\mathbf{U}^{\mathrm{H}}$,
where $\boldsymbol{\Upsilon}$ is the eigenvalue matrix and $\mathbf{U}$
is a square matrix whose columns are the eigenvectors of $\mathbf{A}\mathbf{A}^{\mathrm{H}}$.
Next we construct a matrix $\boldsymbol{\Gamma}\in\mathbb{C}^{\tau_{1}\times M}$
with orthogonal rows, i.e., $\boldsymbol{\Gamma}\boldsymbol{\Gamma}^{\mathrm{H}}=\mathbf{I}_{\tau_{1}}$;
for example, we can select $\boldsymbol{\Gamma}=[\mathbf{I}_{\tau_{1}}\mathbf{0}]$.
Then, Problem (\ref{eq:kk-1}) becomes 
\begin{align}
\min_{\mathbf{E}_{1}} & ||\mathbf{E}_{1}^{\mathrm{H}}\mathbf{U}\boldsymbol{\Upsilon}^{\frac{1}{2}}-\boldsymbol{\Gamma}||_{F}^{2}\nonumber \\
\mathrm{s.t.} & |[\mathbf{E}_{1}]_{m,n}|=1,1\leq m\leq M,1\leq n\leq\tau_{1}.\label{eq:kk-4}
\end{align}
The unconstrained LS solution of Problem (\ref{eq:kk-4}) is $\mathbf{E}_{1}^{\mathrm{LS}}=(\boldsymbol{\Gamma}\boldsymbol{\Upsilon}^{-\frac{1}{2}}\mathbf{U}^{\mathrm{H}})^{\mathrm{H}}$.
By mapping $\mathbf{E}_{1}^{\mathrm{LS}}$ to the unit-modulus constraint,
the final solution to Problem (\ref{eq:kk-4}) is given by 
\begin{equation}
\mathbf{E}_{1}=\exp\left(\mathrm{i}\angle(\boldsymbol{\Gamma}\boldsymbol{\Upsilon}^{-\frac{1}{2}}\mathbf{U}^{\mathrm{H}})^{\mathrm{H}}\right).\label{eq:vf}
\end{equation}

For the design of $\mathbf{E}_{k}$ $(2\leq k\leq K)$, it is straightforward
to formulate a problem similar to (\ref{eq:kk-4}) as follows: \begin{subequations}
\label{hc-a-2} 
\begin{align}
\min_{\mathbf{E}_{k}} & ||\mathbf{Z}_{k}\mathbf{U}\boldsymbol{\Upsilon}^{\frac{1}{2}}-\boldsymbol{\Gamma}||_{F}^{2}\label{eq:obj-1}\\
\mathrm{s.t.} & |[\mathbf{E}_{k}]_{m,n}|=1,1\leq m\leq M,1\leq n\leq\tau_{k}.\label{eq:kk-5}
\end{align}
\end{subequations}Due to the structure of $\mathbf{Z}_{k}$ in (\ref{eq:zz}),
$\boldsymbol{\Gamma}$ should be carefully constructed for better
performance in solving (\ref{hc-a-2}). Here, we propose an AO method
to alternately design $\boldsymbol{\Gamma}$ and $\mathbf{E}_{k}$.

In particular, with a pre-designed $\mathbf{E}_{k}$ derived from
a DFT matrix, $\boldsymbol{\Gamma}$ can be constructed by solving
the following problem 
\begin{equation}
\boldsymbol{\Gamma}=\mathrm{arg}\min_{\boldsymbol{\Gamma}\boldsymbol{\Gamma}^{\mathrm{H}}=\mathbf{I}_{\tau_{k}}}||\mathbf{Z}_{k}\mathbf{U}\boldsymbol{\Upsilon}^{\frac{1}{2}}-\boldsymbol{\Gamma}||_{F}^{2}.\label{fd}
\end{equation}
Problem (\ref{fd}) is an orthogonal Procrustes problem \cite{Xinda2017}.
Define the singular value decomposition of $\mathbf{Z}_{k}\mathbf{U}\boldsymbol{\Upsilon}^{\frac{1}{2}}=\mathbf{P}\boldsymbol{\Xi}\mathbf{Q}^{\mathrm{H}}$,
where $\boldsymbol{\Xi}\in\mathbb{C}^{\tau_{k}L\times M}$ is a diagonal
matrix whose diagonal elements are the singular values of $\mathbf{Z}_{k}\mathbf{U}\boldsymbol{\Upsilon}^{\frac{1}{2}}$,
$\mathbf{P}\in\mathbb{C}^{\tau_{k}L\times\tau_{k}L}$ and $\mathbf{Q}\in\mathbb{C}^{M\times M}$
are unitary matrices. Then, the optimal solution to Problem (\ref{fd})
is given by $\boldsymbol{\Gamma}=\mathbf{P}\mathbf{Q}_{(:,1:\tau_{k}L)}^{\mathrm{H}}$
\cite{Xinda2017}.

The complicated structure of (\ref{eq:zz}) does not lead to a direct
solution for the design of $\mathbf{E}_{k}$. To address this difficulty,
we reconstruct (\ref{eq:obj-1}) via several mathematical transformations
so that $\mathbf{E}_{k}$ can be written in quadratic form. In particular,
denote $\boldsymbol{\Gamma}=[\boldsymbol{\Gamma}_{1}^{\mathrm{T}},\ldots,\boldsymbol{\Gamma}_{\tau_{k}}^{\mathrm{T}}]^{\mathrm{T}}$,
where $\boldsymbol{\Gamma}_{t}\in\mathbb{C}^{L\times M}$ for $1\leq t\leq\tau_{k}$.
With the determined $\boldsymbol{\Gamma}$ and (\ref{eq:zz}), (\ref{eq:obj-1})
is equivalent to 
\begin{align}
 & \sum_{t=1}^{\tau_{k}}||\boldsymbol{\Lambda}_{\mathrm{c}}\mathbf{A}_{\mathrm{c}}^{\mathrm{H}}\mathrm{Diag}\left(\mathbf{e}_{t}\right)\mathbf{U}\boldsymbol{\Upsilon}^{\frac{1}{2}}-\boldsymbol{\Gamma}_{t}||_{F}^{2}\nonumber \\
\stackrel{(a)}{=} & \sum_{t=1}^{\tau_{k}}||\mathbf{T}\mathbf{e}_{t}-\mathrm{vec}\left(\boldsymbol{\Gamma}_{t}\right)||_{2}^{2},\label{eq:ko}
\end{align}
where $\mathbf{T}=(\mathbf{U}\boldsymbol{\Upsilon}^{\frac{1}{2}})^{\mathrm{T}}\odot\boldsymbol{\Lambda}_{\mathrm{c}}\mathbf{A}_{\mathrm{c}}^{\mathrm{H}}$.
Step (a) is due to the property $\mathrm{vec}\left(\mathbf{X}\mathrm{Diag}\left(\mathbf{e}_{t}\right)\mathbf{Y}^{\mathrm{T}}\right)=\left(\mathbf{Y}\odot\mathbf{X}\right)\mathbf{e}_{t}$
\cite{TSP-tensor}. By parallel stacking $\mathbf{F}=[\mathrm{vec}\left(\boldsymbol{\Gamma}_{1}\right),\ldots,\mathrm{vec}\left(\boldsymbol{\Gamma}_{\tau_{k}}\right)]$,
(\ref{eq:ko}) is further equivalent to $||\mathbf{T}\mathbf{E}_{k}-\mathbf{F}||_{F}^{2}$.
Therefore, Problem (\ref{hc-a-2}) is reformulated as 
\begin{align}
\min_{\mathbf{E}_{k}} & ||\mathbf{T}\mathbf{E}_{k}-\mathbf{F}||_{F}^{2}\nonumber \\
\mathrm{s.t.} & |[\mathbf{E}_{k}]_{m,n}|=1,1\leq m\leq M,1\leq n\leq\tau_{k}.\label{eq:kk-5-1}
\end{align}

The unconstrained LS solution of Problem (\ref{eq:kk-5-1}) is $\mathbf{E}_{k}^{\mathrm{LS}}=(\mathbf{T}^{\mathrm{H}}\mathbf{T})^{-1}\mathbf{T}^{\mathrm{H}}\mathbf{F}$.
By mapping $\mathbf{E}_{k}^{\mathrm{LS}}$ to the unit-modulus constraint,
the final solution to Problem (\ref{eq:kk-5-1}) is given by 
\begin{equation}
\mathbf{E}_{k}=\exp\left(\mathrm{i}\angle((\mathbf{T}^{\mathrm{H}}\mathbf{T})^{-1}\mathbf{T}^{\mathrm{H}}\mathbf{F})\right).\label{eq:vf-1}
\end{equation}

\begin{figure}
\centering \includegraphics[width=3in,height=0.7in]{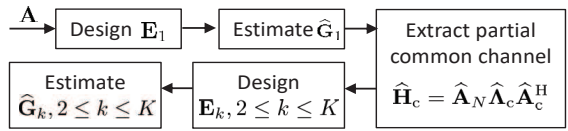}
\caption{Flow chart of training reflection matrix design and cascaded channel
estimation.}
\label{flow} 
\end{figure}

Problem (\ref{fd}) and Problem (\ref{eq:kk-5-1}) are optimized alternately
until a stopping criterion is satisfied. Fig. \ref{flow} shows the
flow chart of the training phase shift matrix design and the cascaded
channel estimation.

\section{Analysis of pilot overhead and computational complexity}

In this section, we analyze the pilot overhead and the computational
complexity of our proposed channel estimation method. We also compare
our results with the other existing algorithms summarized in Table
I. In this section we assume $J_{1}=J_{2}=\cdots=J_{K}=J$ for simplicity.

\begin{table*}
\centering \label{table} \caption{Pilot Overhead and Complexity Comparison of Different Estimation Algorithms.}
\begin{tabular}{|c|c|c|c|}
\hline 
\multicolumn{2}{|c|}{\textbf{Algorithm}} & \textbf{Pilot Overhead}  & \textbf{Complexity}\tabularnewline
\hline 
\multirow{2}{*}{Proposed algorithm} & First coherence block  & $8J-2+(K-1)\left\lceil (8J-2)/L\right\rceil $  & $\mathcal{O}(Ng+8DJ+8KDJ^{4})$\tabularnewline
\cline{2-4} \cline{3-4} \cline{4-4} 
 & Remaining coherence blocks  & $JK$  & $\mathcal{O}(KJ^{3})$\tabularnewline
\hline 
\multicolumn{2}{|c|}{Conventional-OMP algorithm \cite{ris-omp-1}} & $K\left\lceil (8J-2)/N\right\rceil $  & $\mathcal{O}(8KMJD^{2}J^{4}L^{3})$\tabularnewline
\hline 
\multicolumn{2}{|c|}{CS-based algorithm \cite{ris-omp-2}} & $K\left\lceil M/(JL)\right\rceil $  & $\mathcal{O}(N^{3}+D^{3}+KM^{3})$\tabularnewline
\hline 
\multicolumn{2}{|c|}{DS-OMP algorithm \cite{ris-omp-3}} & $K(8J-2)$  & $\mathcal{O}(KND+KMDLJ^{3})$\tabularnewline
\hline 
\end{tabular}
\end{table*}

\subsection{Pilot Overhead}

In the first coherence block, all users need to estimate the full
CSI. The theoretical minimum pilot overhead of user 1 is $\tau_{1}=8J-2$,
and that of user $k,2\leq k\leq K,$ is $\tau_{k}=\left\lceil (8J-2)/L\right\rceil $.
Therefore, the total pilot overhead is $8J-2+(K-1)\left\lceil (8J-2)/L\right\rceil $.
In the remaining channel coherence blocks, each user needs to transmit
$\tau_{k}=J$, $1\leq k\leq K,$ pilots for the estimation of the
cascaded channel gains. Thus, the total pilot overhead in these coherence
blocks is $JK$.

Compared with the existing estimation algorithms in Table I, the proposed
algorithm has a very low pilot overhead for estimating the full CSI
in the first coherence block. When the angle information of the cascaded
channel is estimated, the pilot overhead is further reduced for the
re-estimated cascaded gains in the remaining coherence blocks.

\subsection{Complexity analysis}

We first calculate the computational complexity of Algorithm \ref{Algorithm-G-1}
for user 1. The complexity of Stage 1 in Algorithm \ref{Algorithm-G-1}
mainly stems from the angle rotation operation (\ref{eq:rotation})
which has complexity order of $\mathcal{O}(Ng)$, where $g$ denotes
the number of grid points in the interval $[-\frac{\pi}{N},\frac{\pi}{N}]$.
For a very large $N$, a small value of $g$ is good enough for high
accuracy and low complexity. The complexity of the OMP algorithm is
given by $\mathcal{O}(nml^{3})$, where $n$ is the length of the
measurement data, $m$ is the length of the sparse signal with sparsity
level $l$ \cite{mmwave-omp}. Thus, the complexity of the OMP algorithm
in Stage 2 is $\mathcal{O}(8DJ^{4})$. Stage 3 can be regarded as
an OMP with one sparse signal, thus its complexity is on the order
of $\mathcal{O}(8DJ)$. Therefore, the estimation complexity for user
1 is $\mathcal{O}(Ng+8DJ+8DJ^{4})$. The computational complexity
for user $k,2\leq k\leq K$, is due to the use of OMP for solving
Problem (\ref{eq:zc}), and this estimation complexity for user $k,2\leq k\leq K,$
is $\mathcal{O}(8DJ^{4})$. Therefore, the total estimation complexity
for $K$ users in the first coherence block is given by $\mathcal{O}(Ng+8DJ+K8DJ^{4})$.

In the remaining coherence blocks, only cascaded channel gains need
to be updated by using the LS solutions in (\ref{eq:Y1-1}), the computational
complexity of which is on the order of $\mathcal{O}(J^{3})$. Therefore,
the total estimation complexity for $K$ users in these coherence
blocks is on the order of $\mathcal{O}(KJ^{3})$.

Since $L\ll N(M)$, $J\ll N(M)$ and $g\ll N$, the complexity of
the proposed algorithm in every coherence block is much lower than
the other estimation algorithms in the existing literature, as shown
in Table I.

\section{Simulation Results}

In this section, we present extensive simulation results to validate
the effectiveness of the proposed channel estimation method. All results
are obtained by averaging over 500 channel realizations. The uplink
carrier frequency is set as $f_{c}=28$ GHz. The channel complex gains
are generated according to $\alpha_{l}\sim\mathcal{CN}(0,10^{-3}d_{\mathrm{BR}}^{-2.2})$
and $\beta_{k,j}\sim\mathcal{CN}(0,10^{-3}d_{\mathrm{RU}}^{-2.8})$,
where $d_{\mathrm{BR}}$ represents the distance from the BS to the
RIS and is assumed to be $d_{\mathrm{BR}}=100$ m, while $d_{\mathrm{RU}}$
denotes the distance between the RIS and users and is set as $d_{\mathrm{RU}}=10$
m. The SNR is defined as $\mathrm{SNR}=10\log(10^{-6}d_{\mathrm{BR}}^{-2.2}d_{\mathrm{RU}}^{-2.8}p/\delta^{2})$,
and the transmit power for all users is set as $p=1$ W. The angles
$\{\phi_{l},\theta_{l},\vartheta_{k,j}\}$ are continuous and uniformly
distributed over $[0,\pi)$. The number of users is $K=4$. Unless
otherwise noted, the number of paths in the mmWave channels is equal
to 4 according to the experimental measurements in dense urban environments
reported in \cite{mmWave-channel}, thus the number of paths in the
cascaded channel are set as $L=5$ and $J_{1}=\cdots=J_{K}=4$. The
antenna element space at the BIS and RIS are set as $d_{\mathrm{BS}}=\frac{\lambda_{c}}{2}$
and $d_{\mathrm{RIS}}=\frac{\lambda_{c}}{4}$, respectively. The normalized
mean square error (NMSE) of the cascaded channel matrix is defined
as 
\begin{align*}
\mathrm{NMSE} & =\mathbb{E}\{||\widehat{\mathbf{G}}_{k}-\mathbf{G}_{k}||_{F}^{2}\}/\mathbb{E}\{||\mathbf{G}_{k}||_{F}^{2}\}.
\end{align*}

\begin{figure}
\centering \includegraphics[width=3.5in,height=2.5in]{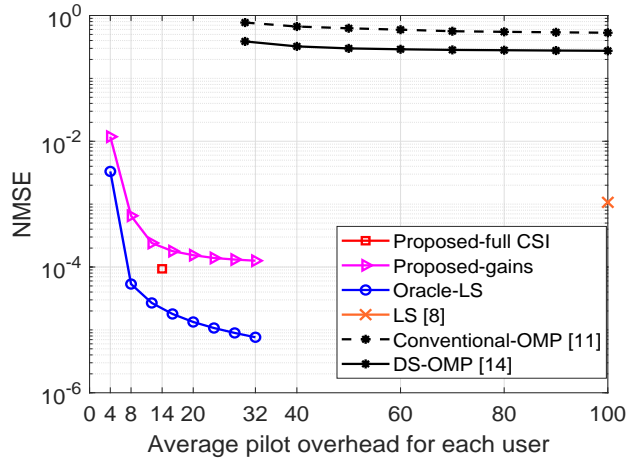}
\caption{NMSE versus pilot overhead, when $N=100$, $M=100$, $L=5$, $J=4$
and SNR=$0$ dB.}
\label{pilot} 
\end{figure}

The estimation algorithms considered in the simulations are as follows: 
\begin{itemize}
\item Proposed-full-CSI: The channels are estimated using the proposed DFT-OMP-based
algorithm in Algorithm \ref{Algorithm-G-1} in the first coherence
block. 
\item Proposed-gains: When the angle information estimated in the first
coherence block is fixed, the channels are determined by only estimating
the cascaded channel gains via the LS method in (\ref{eq:Y1-1}). 
\item Oracle-LS: The angle information is perfectly known at the BS, and
the cascaded channel gains are estimated by (\ref{eq:Y1-1}). This
algorithm can be regarded as the performance upper bound. 
\item LS \cite{LS-mvue}: The channels are estimated using the LS estimator
(\ref{ls}) with the optimal training phase shift matrix drawn from
a DFT matrix. 
\item Conventional-OMP \cite{ris-omp-1}: After approximating the cascaded
channel using the VAD representations in (\ref{ds}), a sparse signal
reconstruction problem is constructed by vectorizing the measurement
matrix. Then, the cascaded channels are estimated directly using OMP. 
\item DS-OMP \cite{ris-omp-3}: The double-sparse structure of the angular
domain sparse cascaded channel matrix $\mathbf{X}_{k}$ in (\ref{ds})
is exploited. The cascaded channels are estimated using OMP for each
non-zero row of $\mathbf{X}_{k}$. 
\end{itemize}
Fig. \ref{pilot} illustrates the impact of pilot overhead on the
estimation performance when the SNR is $0$ dB. Since the number of
time slots allocated to each user for channel estimation in the Proposed-full-CSI
algorithm is different, we choose the average number of time slots
for each user as the x-axis measurement, denoted as $T$. It is obvious
that a larger pilot overhead leads to better NMSE performance for
all channel estimation algorithms. The Proposed-full-CSI algorithm
with $T=14$ time slots outperforms the LS algorithm with $T=M=100$
time slots. This is because the cascaded channel estimated by the
Proposed-full-CSI algorithm exploits the low-rank characteristic of
the mmWave channel, while the LS algorithm ignores the channel sparsity.
When the angle information is estimated, the Proposed-gains algorithm
only needs $T=2J=8$ time slots to surpass the performance of the
LS algorithm. In addition, we observe that even though the two OMP-based
algorithms in \cite{ris-omp-1} and \cite{ris-omp-3} use many more
pilots than the theoretical minimum pilot overhead shown in Table
I, they are unable to achieve good estimation performance. This is
because the algorithm in \cite{ris-omp-1} completely ignores the
double sparse structure of the cascaded channels, resulting in many
false alarm estimates. The algorithm in \cite{ris-omp-3} ignores
the impact of power leakage and ideally assumes that the number of
multipaths is known, resulting in the real low-power paths being replaced
by virtual high-power paths. The impact of power leakage is addressed
in the proposed estimation algorithm by using the angle rotation operation,
designing the optimal phase shift matrix, and enlarging the dimension
of the dictionary. Finally, the proposed channel estimation strategy
of using the Proposed-full-CSI algorithm followed by Proposed-gains,
can achieve significantly improved estimation performance with very
little pilot overhead, compared with the existing channel estimation
algorithms.

\begin{figure}
\centering \includegraphics[width=3.5in,height=2.5in]{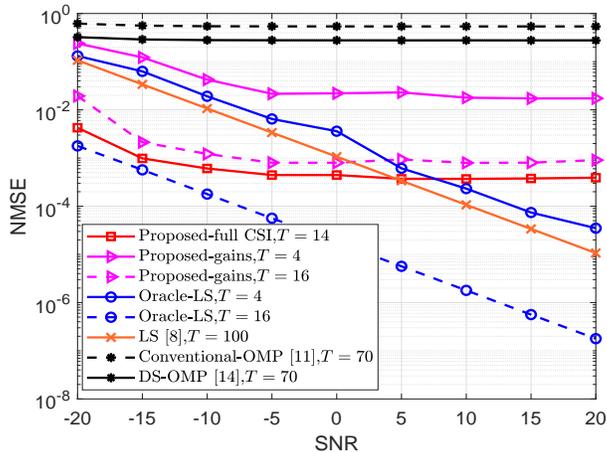}
\caption{NMSE versus SNR, when $N=100$, $M=100$, $L=5$ and $J=4$.}
\label{SNR} 
\end{figure}

Fig. \ref{SNR} displays NMSE performance as a function of SNR for
difference channel estimation methods. At low SNR, it can be seen
that the performance of the proposed algorithms are better than that
of LS. As SNR increases, the estimation accuracy of the proposed algorithms
increases but will reach saturation at relatively high SNR. The reasons
for the error floor are twofold: one is the slight AoA steering matrix
non-orthogonality since $N$ is finite, the other is the mismatch
between the estimated cascaded AoD cosines and the real cascaded AoD
cosines due to the fact that OMP selects the estimation angles from
the discrete grid.

\begin{figure}
\centering \includegraphics[width=3.5in,height=2.5in]{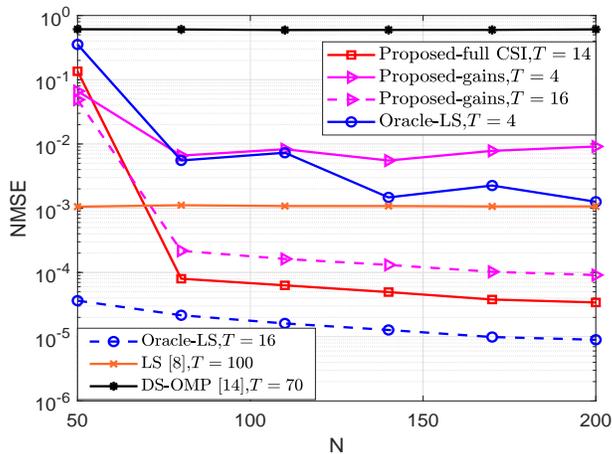} \caption{NMSE versus the number of antennas, when $M=100$, $L=5$, $J=4$
and SNR=$0$ dB.}
\label{N} 
\end{figure}

We next show the NMSE performance with various numbers of antennas
$N$ when SNR=$0$ dB in Fig. \ref{N}. From the figure, when $N$
increases, the performance of the LS method remains stable and is
not affected by the size of the estimated channel, because there are
enough time slots to support LS estimation in the spatial domain.
The OMP-based benchmark consistently performs poorly due to its serious
power leakage effect. On the other hand, when $N$ is larger than
80, the Proposed-full-CSI algorithm works well, because the resolution
of the DFT in Algorithm \ref{Algorithm-dft} improves with larger
$N$. At the same time, the angle rotation operation in Algorithm
\ref{Algorithm-dft} can also alleviate the impact of power leakage.
In addition, when the pilot overhead increases from $T=J$ to $T=4J$
(i.e., from 4 to 16), the performance of the Proposed-gains algorithm
improves since more pilot overhead can provide more measurement data
diversity for the algorithm.

\begin{figure}
\centering \includegraphics[width=3.5in,height=2.5in]{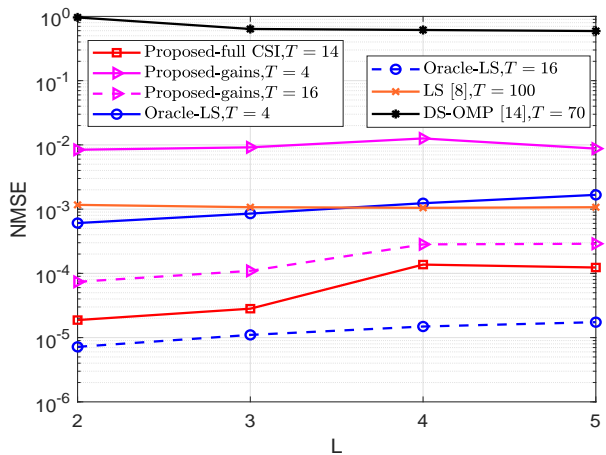} \caption{NMSE versus the number of paths from the BS to the RIS $L$, when
$N=100$, $M=100$, $J=4$ and SNR=$0$ dB.}
\label{L} 
\end{figure}

Fig. \ref{L} shows the impact of the number of spatial paths between
the BS and the RIS. It is clear that the number of spatial paths has
no effect on the LS method. However, the performance of the proposed
algorithms degrades when the number of spatial paths increases, due
to the fact that the number of parameters (sparsity level) to be estimated
increases.

\section{Conclusions}

In this paper, we developed a cascaded channel estimation method for
RIS-aided uplink multiuser mmWave systems with much less pilot overhead.
Our algorithm takes advantage of angle information that remains essentially
static for many coherence blocks, exploits the linear correlation
among cascaded paths, as well as the reparameterized CSI of the common
BS-RIS channel. The theoretical minimum pilot overhead was characterized,
and training reflection matrices were designed. Simulation results
showed that the NMSE performance of the proposed algorithm outperforms
the existing OMP-based algorithms and the pilot overhead required
by the proposed algorithm is much less than for existing methods..

\appendices{}

\section{The proof of Lemma \ref{othogonal}\label{subsec:The-proof-of-3}}

We calculate 
\begin{align}
\mathbf{a}_{N}^{\mathrm{H}}(\psi_{l})\mathbf{a}_{N}(\psi_{i})= & \sum_{m=1}^{N}e^{-\mathrm{i}2\pi(m-1)(\psi_{i}-\psi_{l})}\nonumber \\
= & \frac{1-e^{-\mathrm{i}2\pi N(\psi_{i}-\psi_{l})}}{1-e^{-\mathrm{i}2\pi(\psi_{i}-\psi_{l})}}.\label{eq:Phi-1-1}
\end{align}
The product $\mathbf{a}_{N}^{\mathrm{H}}(\psi_{l})\mathbf{a}_{N}(\psi_{i})$
is bounded for any $l\neq i$ as $N\rightarrow\infty$ and thus $\lim_{N\rightarrow\infty}\frac{1}{N}\mathbf{a}_{N}^{\mathrm{H}}(\psi_{l})\mathbf{a}_{N}(\psi_{i})=0$.
When $l=i$, direct calculation yields that $\mathbf{a}_{N}^{\mathrm{H}}(\psi_{l})\mathbf{a}_{N}(\psi_{j})=N$
and hence $\lim_{N\rightarrow\infty}\frac{1}{N}\mathbf{a}_{N}^{\mathrm{H}}(\psi_{l})\mathbf{a}_{N}(\psi_{i})=1$.
Therefore, when $N\rightarrow\infty$, the limit of (\ref{eq:Phi-1-1})
is 
\begin{align}
 & \lim_{N\rightarrow\infty}\frac{1}{N}\mathbf{a}_{N}^{\mathrm{H}}(\psi_{l})\mathbf{a}_{N}(\psi_{i})=\delta\left(\psi_{i}-\psi_{l}\right),\label{eq:vd-1}
\end{align}
where $\delta(\cdot)$ is the Dirac delta function.

The proof is completed.

\section{The proof of Lemma \ref{DFT}\label{subsec:The-proof-of-1}}

Let us first consider the case $\psi_{l}\in[0,\frac{d_{\mathrm{BS}}}{\lambda_{c}})$.
Then, the $(n,l)$-th element of $\mathbf{U}_{N}^{\mathrm{H}}\mathbf{A}_{N}$
is calculated in 
\begin{align}
  [\mathbf{U}_{N}^{\mathrm{H}}\mathbf{A}_{N}]_{n,l}=&\left[\mathbf{U}_{N}^{\mathrm{H}}\mathbf{a}_{N}(\psi_{l})\right]_{n}\nonumber \\
= & \sqrt{\frac{1}{N}}\sum_{m=1}^{N}e^{\mathrm{i}\frac{2\pi}{N}(n-1)(m-1)}e^{-\mathrm{i}2\pi(m-1)\psi_{l}}\nonumber \\
= & \sqrt{\frac{1}{N}}\sum_{m=1}^{N}e^{-\mathrm{i}2\pi(m-1)(\psi_{l}-\frac{n-1}{N})}\nonumber \\
= & \sqrt{\frac{1}{N}}\frac{1-e^{\mathrm{i}2\pi N(\frac{n-1}{N}-\psi_{l})}}{1-e^{\mathrm{i}2\pi(\frac{n-1}{N}-\psi_{l})}}.\label{eq:Phi-1}
\end{align}
According to the proof in Appendix \ref{subsec:The-proof-of-3}, when
$N\rightarrow\infty$, the limit of (\ref{eq:Phi-1}) is 
\begin{align}
 & \lim_{N\rightarrow\infty}\left|[\mathbf{U}_{N}^{\mathrm{H}}\mathbf{a}_{N}(\psi_{l})]_{n}\right|=\sqrt{N}\delta\left(\frac{n-1}{N}-\psi_{l}\right).\label{eq:vd}
\end{align}
Hence, there always exist some integers $n_{l}=N\psi_{l}+1$ such
that $|[\mathbf{U}_{N}^{\mathrm{H}}\mathbf{a}_{N}(\psi_{l})]_{n_{l}}|=\sqrt{N}$,
and the other elements of $\mathbf{U}_{N}^{\mathrm{H}}\mathbf{a}_{N}(\psi_{l})$
are zero. In other words, $\mathbf{U}_{N}^{\mathrm{H}}\mathbf{A}_{N}$
is a sparse matrix with all powers being concentrated on the points
$(n_{l},l),$ $\forall l$.

When $\psi_{l}\in[-\frac{d_{\mathrm{BS}}}{\lambda_{c}},0)$, using
the fact that $e^{\mathrm{i}x}=e^{\mathrm{i}(x+2\pi)}$, (\ref{eq:Phi-1})
is equivalent to 
\begin{align}
  [\mathbf{U}_{N}^{\mathrm{H}}\mathbf{A}_{N}]_{n,l}=&\left[\mathbf{U}_{N}^{\mathrm{H}}\mathbf{a}_{N}(\psi_{l})\right]_{n}\nonumber \\
= & \sqrt{\frac{1}{N}}\sum_{m=1}^{N}e^{-\mathrm{i}\left[2\pi(m-1)(\psi_{l}-\frac{n-1}{N})+2\pi(m-1)\right]}\nonumber \\
= & \sqrt{\frac{1}{N}}\sum_{m=1}^{N}e^{-\mathrm{i}2\pi(m-1)(\psi_{l}-\frac{n-1}{N}+1)}\label{eq:ddd}
\end{align}
When $N\rightarrow\infty$, the limit of (\ref{eq:ddd}) is 
\begin{align}
\lim_{N\rightarrow\infty}\left|[\mathbf{U}_{N}^{\mathrm{H}}\mathbf{a}_{N}(\psi_{l})]_{n}\right|=\sqrt{N}\delta\left(\psi_{l}-\frac{n-1}{N}+1\right).\label{fe}
\end{align}

Hence, there always exist some integers $n_{l}=N+N\psi_{l}+1$ such
that $|[\mathbf{U}_{N}^{\mathrm{H}}\mathbf{a}_{N}(\psi_{l})]_{n}|=\sqrt{N}$,
and the other elements of $[\mathbf{U}_{N}^{\mathrm{H}}\mathbf{a}_{N}(\psi_{l})]_{n}$
are zero. Combining (\ref{eq:vd}) and (\ref{fe}), we arrive at (\ref{eq:ss-1}).

The proof is completed.

 \bibliographystyle{IEEEtran}
\bibliography{bibfile}

\end{document}